\documentclass[iop]{emulateapj}

\usepackage{apjfonts}
\usepackage{graphicx}
\usepackage{epstopdf}
\usepackage{ulem}
\usepackage[utf8]{inputenc}
\usepackage[T1]{fontenc}
\usepackage{url}
\usepackage{hyperref}
\usepackage{mdwlist}
\usepackage{amsmath} 
\usepackage{enumitem}

\makeatletter
\def\url@leostyle{%
 \@ifundefined{selectfont}{\def\UrlFont{\sf}}{\def\UrlFont{\small\ttfamily}}}
\makeatother
\begin{document}

\newcommand{\ls}{{_<\atop^{\sim}}}
\newcommand{\gs}{{_>\atop^{\sim}}}
\def \spose#1{\hbox  to 0pt{#1\hss}}  
\def \ls{\mathrel{\spose{\lower 3pt\hbox{$\sim$}}\raise  2.0pt\hbox{$<$}}}
\def \gs{\mathrel{\spose{\lower  3pt\hbox{$\sim$}}\raise 2.0pt\hbox{$>$}}}
\newcommand{\Ha}{\hbox{{\rm H}$\alpha$}}
\newcommand{\Hb}{\hbox{{\rm H}$\beta$}}
\newcommand{\Ovi}{\hbox{{\rm O}\kern 0.1em{\sc vi}}}
\newcommand{\OIII}{\hbox{[{\rm O}\kern 0.1em{\sc iii}]}}
\newcommand{\OII}{\hbox{[{\rm O}\kern 0.1em{\sc ii}]}}
\newcommand{\NII}{\hbox{[{\rm N}\kern 0.1em{\sc ii}]}}
\newcommand{\SII}{\hbox{[{\rm S}\kern 0.1em{\sc ii}]}}
\newcommand{\angstrom}{\textup{\AA}}
\newcommand\ionn[2]{#1$\;${\scshape{#2}}}

\font\btt=rm-lmtk10


\title{Detecting Radio-AGN signatures in Red geysers}

\shorttitle{Detecting Radio-AGN signatures in Red geysers}

\shortauthors{Roy et al.}


\author{Namrata Roy\altaffilmark{1}\dag, Kevin Bundy\altaffilmark{1,2}, Edmond Cheung\altaffilmark{3}, Wiphu Rujopakarn\altaffilmark{3,4}, Michele Cappellari\altaffilmark{5}, Francesco Belfiore\altaffilmark{1}, Renbin Yan\altaffilmark{6}, Tim Heckman\altaffilmark{7}, Matthew Bershady\altaffilmark{8}, Jenny Greene\altaffilmark{9}, Kyle Westfall\altaffilmark{1,2}, Niv Drory\altaffilmark{10}, Kate Rubin\altaffilmark{11}, David Law\altaffilmark{12}, Kai Zhang\altaffilmark{6}, Joseph Gelfand\altaffilmark{13,14}, Dmitry Bizyaev\altaffilmark{15,16}, David Wake\altaffilmark{8,17}, Karen Masters\altaffilmark{18}, Daniel Thomas\altaffilmark{18}, Cheng Li\altaffilmark{3}, Rogemar A. Riffel\altaffilmark{19,20}}
\altaffiltext{1} {Department of Astronomy and Astrophysics, University of California, 1156 High Street, Santa Cruz, CA 95064}
\altaffiltext{2}{UCO/Lick Observatory, Department of Astronomy and Astrophysics, University of California, 1156 High Street, Santa Cruz, CA 95064}
\altaffiltext{3}{Kavli Institute for the Physics and Mathematics of the Universe (WPI), The University of Tokyo Institutes for Advanced Study, The University of Tokyo, Kashiwa, Chiba 277-8583, Japan}
\altaffiltext{4}{Department of Physics, Faculty of Science, Chulalongkorn University, 254 Phayathai Road, Pathumwan, Bangkok 10330, Thailand}
\altaffiltext{5}{Sub-department of Astrophysics, University of Oxford, Denys Wilkinson Building, Keble Road, Oxford OX1 3RH}
\altaffiltext{6}{Department of Physics and Astronomy, University of Kentucky, 505 Rose Street, Lexington, KY 40506-0055, USA}
\altaffiltext{7}{Center for Astrophysical Sciences, Department of Physics \& Astronomy, The Johns Hopkins University, Baltimore, Maryland 21218}
\altaffiltext{8}{Department of Astronomy, University of Wisconsin-Madison, 475 North Charter Street, Madison, WI 53706, USA}
\altaffiltext{9}{Department of Astrophysical Sciences, Princeton University, Princeton, NJ 08544, USA}
\altaffiltext{10}{McDonald Observatory, Department of Astronomy, University of Texas at Austin, 1 University Station, Austin, TX 78712-0259, USA}
\altaffiltext{11}{Harvard-Smithsonian Center for Astrophysics, 60 Garden Street, Cambridge, MA 02138, USA}
\altaffiltext{12}{Space Telescope Science Institute, 3700 San Martin Drive, Baltimore, MD 21218, USA}
\altaffiltext{13}{NYU Abu Dhabi, P.O. Box 129188, Abu Dhabi, UAE}
\altaffiltext{14}{Center for Cosmology and Particle Physics, New York University, Meyer Hall of Physics, 4 Washington Place, New York, NY 10003, USA}
\altaffiltext{15}{Apache Point Observatory and New Mexico State University, P.O. Box 59, Sunspot, NM, 88349-0059, USA}
\altaffiltext{16}{Sternberg Astronomical Institute, Moscow State University, Moscow}
\altaffiltext{17}{Department of Physical Sciences, The Open University, Milton Keynes, MK7 6AA, UK}
\altaffiltext{18}{Institute for Cosmology and Gravitation, University of Portsmouth, Dennis Sciama Building, Burnaby Road, Portsmouth PO1 3FX}
\altaffiltext{19}{Universidade Federal de Santa Maria, Departamento de F\'\i sica, CCNE, 97105-900, Santa Maria, RS, Brazil}
\altaffiltext{20}{Laborat\'orio Interinstitucional de e-Astronomia - LIneA,  Rio de Janeiro, RJ, Brazil}

\altaffiltext{\dag}{naroy@ucsc.edu}


\begin{abstract}

A new class of quiescent galaxies harboring possible AGN-driven winds has been discovered using spatially resolved optical spectroscopy from the ongoing SDSS-IV MaNGA survey. These galaxies, termed ``red geysers'', constitute $5-10\%$ of the local quiescent population and are characterized by narrow bisymmetric patterns in ionized gas emission features. Cheung et al. argued that these galaxies host large-scale AGN-driven winds that may play a role in suppressing star formation at late times.  In this work, we test the hypothesis that AGN activity is ultimately responsible for the red geyser phenomenon.  We compare the nuclear radio activity of the red geysers to a matched control sample with similar stellar mass, redshift, rest frame $NUV-r$ color, axis ratio and presence of ionized gas. We have used the 1.4 GHz radio continuum data from VLA FIRST survey to stack the radio flux from the red geyser and control samples. In addition to a 3 times higher FIRST detection rate, we find that red geysers have a 5$\sigma$ higher level of average radio flux than control galaxies. After restricting to rest-frame $NUV-r$ color $>$ 5 and checking mid-IR WISE photometry, we rule out star formation contamination and conclude that red geysers are associated with more active AGN. Red geysers and a possibly-related class with disturbed H$\alpha$ emission account for 40\% of all radio-detected red galaxies with $\rm log~(M_\star/M_\odot) < 11$. Our results support a picture in which episodic AGN activity drives large-scale-relatively weak ionized winds that may provide a feedback mechanism for many early-type galaxies.

\end{abstract}



\section{Introduction} \label{sec:Introduction}

The level of star formation in galaxies is known to be bimodal \citep{blanton03, strateva01, kauffman03}, with star--forming galaxies often referred to as the ``blue cloud'' and galaxies without significant star formation fall under the ``red sequence'' category. The latter is characterized by old stellar populations ($\rm\gs6~Gyr$) and short star-formation timescales ($\rm\ls1~Gyr$; \citealt{tinsley79, worthey92, trager00, thomas05, graves08, conroy14, worthey14, choi14}). The abundance of these quiescent galaxies has increased by several factors since $\rm z\sim2$ \citep{bell04, bundy06, faber07, ilbert10, moustakas13} which implies that more and more galaxies are transitioning to quiescence. The increase in the red-and-dead population indicates that once galaxies shut off their star formation by some mechanism, they must stay quenched for a long time.

A permanent shut down of star formation is hard to explain, because the quiescent population is not devoid of gas and is also capable of accreting new gas to eventually start forming stars again. Major surveys have shown an abundance of gas in quiescent galaxies \citep{binette94, buson93, demoulin84}, which if left to itself, should ultimately cool and form stars. This gas comes from a variety of sources like stellar mass loss from evolved stars \citep[e.g.,][]{mathews03, ciotti07} or minor mergers. If all this gas formed stars, we would expect the global stellar mass density to be larger by factors of a few than the observed at z = 0. This implies that an additional feedback mechanism is required to maintain the suppression of star formation in galaxies on the red sequence \citep{benson03}.

While a number of feedback mechanisms have been proposed, including interstellar medium (ISM) heating from stellar winds \citep{conroy15} and gravitational effects induced by galaxy bulges \citep{martig09}, the most popular explanation has been active galactic nuclei (AGN) feedback \citep{binney95, ciotti01, croton06, fabian12, yuan14, heckman14}. It states that the energy released from the central AGN in the host galaxy in the form of winds, outflows or jets can significantly effect the evolution of the galaxy by feedback mechanism which can take place via two different ways \citep{fabian12, morganti17}. The ``quasar'' or ``radiative'' mode feedback, mostly associated with luminous AGN or massive quasar, release huge amount of energy to their surroundings by radiation from the accretion disk and drive powerful gas outflows, which remove gas altogether from the galactic potential well \citep{cattaneo09, fabian12}. The ``radio-mode'', on the other hand, is prevalent mostly in low to moderate luminosity AGN where the black hole accretes at a lower rate and the radio AGN present in the center of the galaxy deposits energy into the surrounding gas via jets or winds, heating it and suppressing star formation \citep{binney95, ciotti01, croton06, bower06, ciotti07, ciotti10, mcnamara07, cattaneo09, fabian12, yuan14, heckman14}. Direct observational evidence for this low-luminosity radio-mode or ``maintenance mode'' feedback is limited to several nearby clusters \citep{cattaneo09, dunn06,  fabian94, fabian12, fabian06, mcnamara07}. Evidence for this mechanism in more typical galaxies remain elusive.


Recently, \cite{cheung16} discovered a new class of quiescent galaxies, referred to as ``red geysers'', that show distinctive emission line patterns showing gas outflows from the center and kinematic properties (explained in detail in \S\ref{subsub:redgeys}) which may signal AGN maintenance-mode feedback in action. Based on spatially resolved information from Sloan Digital Sky Survey-IV (SDSS-IV) Mapping Nearby Galaxies at Apache Point Observatory (MaNGA) survey \citep{bundy15}, this class of quiescent galaxies appears to host large scale winds of ionized gas that align with bi-symmetric enhancements in the spatial distribution of strong emission lines like H$\alpha$. Ionized emission extends throughout the entire galaxy with line ratios similar to LIER-like (low ionization emission region) galaxies \citep{belfiore16}. In addition to their enhanced bisymmetric line emission, the red geysers also exhibit gas  kinematics consistent with outflowing winds. The gradient of the gas velocity field aligns with the position angle of the emission pattern, but is largely misaligned with the major or minor axes derived from the stellar velocity field. The gas velocity values can reach $\rm\sim300\ km~s^{-1}$, a value that is difficult to explain by orbital motion from the galaxy's gravitational potential, considering the mass range of the galaxies. 

Early-type galaxies with accreted disks, as studied by \cite{chen16} and \cite{lagos15}, can show similar kinematic features as red geysers, but those features are formed due to a completely different phenomena. The accreted gas coming in  from random directions will gradually align itself with either major or minor axis through gravitational torques generated by the galaxy's potential well. Hence, while a misalignment in the velocity gradient of stars and gas can occur for these galaxies too, often the misalignment angle is 90$^{\circ}$ or 0$^{\circ}$/ 180$^{\circ}$ depending on whether a polar disk or co-rotating/ counter rotating disk is formed. Some galaxies with accreted disks might show similar H$\alpha$ equivalent width (EW) distributions as red geysers. \cite{cheung16} rejected the disk interpretation through detailed Jeans Anisotropic modeling \citep[JAM,][]{cappellari08} of the prototypical red geyser with 99\% confidence which demonstrated that the gas velocity in this source is too high (difference between observed gas velocity and expected velocity from modeling being $\sim \rm 100~km~s^{-1}$) to be described by the orbital motion. Given similar high velocities and other common features shared among all the red geysers, outflowing winds emerge as a compelling interpretation (Bundy et al. in preparation), making the question of whether AGNs are capable of driving these winds particularly important. 

A critical first step is to test the hypothesis that the red geyser population is more likely to host an active AGN compared to quiescent galaxies with similar global galaxy properties. For the prototypical red geyser named ``Akira'', \cite{cheung16} showed that the host galaxy has a weakly and/or radiatively-inefficient supermassive black hole with very low Eddington ratio ($\lambda = 3.9 \times 10^{-4}$), accreting mass from a low-mass companion galaxy. It was detected as a central radio point source. 

The goal of this work is to search for radio-mode AGNs in the entire red geyser sample. We analyze stacked radio flux from Very Large Array (VLA) Faint Images of the Radio Sky at Twenty-Centimeters (FIRST) survey and find a higher value of radio flux from the red geyser candidates than the comparison sample of quiescent galaxies. We have excluded possible star formation contamination and/or galaxies with embedded disks from our sample by using optical and infrared photometry. \S \ref{sec:data} describes the optical, infrared and radio data that we have used in this work. In \S \ref{sub:sample}, we discuss in detail, the red geysers and the control sample chosen from the MaNGA local quiescent population. The technical details of radio aperture photometry and the stacking analysis have been narrated in \S \ref{sub:photometry}. The results thus obtained from the stacked radio flux are described in \S \ref{sec:results}. The implication of the results are given in \S \ref{sec:conclusion}. 

Throughout this paper, we assume a flat cosmological model with $H_{0} = 70$ km s$^{-1}$ Mpc$^{-1}$, $\Omega_{m} = 0.30$, and  $\Omega_{\Lambda} =0.70$, and all magnitudes are given in the AB magnitude system. 

\section{Data} \label{sec:data}

\subsection { The MaNGA survey}

Our sample comes from the ongoing SDSS-IV MaNGA survey \citep{blanton17, bundy15, drory15, law15, yan16, sdss16}. MaNGA is an integral field spectroscopic survey that provides spatially resolved spectroscopy for nearby galaxies ($\rm z\sim0.03$) with an effective spatial resolution of $2.5''$ (full width at half-maximum; FWHM). The MaNGA survey uses the SDSS 2.5 meter telescope in spectroscopic mode \citep{gunn06} and the two dual-channel BOSS spectrographs \citep{smee13} that provide continuous wavelength coverage from the near-UV to the near-IR: $\rm3,600-10,000$ \AA. The spectral resolution varies from $\rm R\sim1400$ at 4000~\AA~ to $\rm R\sim2600$ at 9000~\AA. An $r$-band signal-to-noise $(SN)$ of $\rm 4-8$~\AA$^{-1}$ is achieved in the outskirts (i.e., $\rm1-2~R_{e}$) of target galaxies with an integration time of approximately 3-hr. 
MaNGA will observe roughly 10,000 galaxies with $\rm \log~(M_*/ M_{\odot})\gs9$ across $\sim$ 2700 deg$^{2}$ over its 6~yr duration. In order to balance radial coverge versus spatial resolution, MaNGA observes two thirds of its galaxy sample to $\sim$ 1.5~R$_e$ and one third to 2.5~R$_e$. The MaNGA target selection is described in detail in \cite{wake17}.

The raw data are processed with the MaNGA Data Reduction Pipeline (DRP) \citep{law16}. An individual row-by-row algorithm is used to extract the fiber flux and derive inverse variance spectra from each exposure, which are then wavelength calibrated, flat-fielded and sky subtracted. We use the MaNGA sample and data products drawn from the MaNGA Product Launch-5 (MPL-5) and Data Release 13. The data products are identical to those released as part of the SDSS Data Release 14 \citep[DR14,][]{masters18}. We use spectral measurements and other analyses carried out by a preliminary version of the MaNGA Data Analysis Pipeline (DAP), specifically version 2.0.2.\footnote{This version of the code will be made public in the upcoming SDSS Data Release 15 (DR15; Aguado et al. 2019, {\it submitted}).  An overview of the DAP used for DR15 and its products is described by Westfall et al., {\it in prep}, and assessments of its emission-line fitting approach is described by Belfiore et al., {\it in prep}.}
In brief, the data we use here are based on DAP analysis of each spaxel in the MaNGA datacubes.  The DAP first fits the stellar continuum of each spaxel to determine the stellar kinematics using the Penalised Pixel-fitting algorithm {\tt pPXF} \citep{cappellari04, cappellari17}  and templates based on the MILES stellar library \citep{MILES}.  The templates are a hierarchically clustered distillation of the full MILES stellar library into 49 templates.  This small set of templates provide statistically equivalent
fits to those that use the full library of 985 spectra in the MILES stellar library.  The emission-line regions are masked during this fit.
The DAP then subtracts the result of the stellar continuum modeling to provide a (nearly) continuum-free spectrum that is used to fit the nebular emission lines.  This version of the DAP treated each line independently, fitting each for its flux, Doppler shift, and width, assuming a Gaussian profile shape.  This is different from the approach
used by the DAP for DR15, which is to tie the velocities of all lines to a single value and to impose fixed flux ratios for the [OI], [OIII], and [NII] line doublets.  A detailed comparison of the results from the DR15 and MPL-5 versions of the DAP show that the different approach taken by the latter, and used for our analysis, has a negligible influence on our results.
The final output from the DAP are gas and stellar kinematics, emission line properties and stellar absorption indices.

We use ancillary data drawn from the NASA-Sloan Atlas\footnote{\href{http://www.nsatlas.org}{http://www.nsatlas.org}} (NSA) catalog which reanalyzes images and derives morphological parameters for local galaxies observed in Sloan Digital Sky Survey imaging. It compiles spectroscopic redshifts, UV photometry (from GALEX; \citealt{martin05}), stellar masses, and structural parameters. We have specifically used spectroscopic redshifts and stellar masses from the NSA catalog.

\subsection{ The FIRST survey}

The radio data studied in this paper comes from the VLA Faint Images of the Radio Sky at Twenty Centimeters (FIRST; \citealt{becker95}) survey, which obtained data at frequency channels centered at 1.36 GHz and 1.4 GHz over 10,000 square degrees in the North and South Galactic Caps. The source detection threshold is $\sim1$ mJy, corresponding to a source density of $\rm\sim90~sources~deg^{-2}$. FIRST images have 1.8$''$ pixels with a resolution of $\sim5''$ and typical rms of 0.15~mJy. The astrometric accuracy of each source is $0.5- 1''$ at the source detection threshold. Since FIRST survey area was designed to overlap with the Sloan Digital Sky Survey \citep[SDSS;][]{york00, abazajian09}, most MaNGA targets have FIRST data coverage. For our sample of interest, 93\% of the red geysers have FIRST coverage. However, the 1 mJy threshold results in non detections for most MaNGA galaxies. 

For each pointing center, there are twelve adjacent single-field pointings that are co-added to produce the final FIRST image. Sources are extracted from the co-added reduced images and fit by two dimensional Gaussians to derive peak flux, integrated flux densities and size information. The current FIRST catalog is accessible from the FIRST search page\footnote{\href{http://sundog.stsci.edu/cgi-bin/searchfirst}{http://sundog.stsci.edu/cgi-bin/searchfirst}. The full images are available in \textit{\href{ftp://archive.stsci.edu/pub/vla\_first/data} {ftp://archive.stsci.edu/pub/vla\_first/data}}}.




\subsection { SDSS+WISE star formation rates}

In order to assess possible contamination from obscured star formation, we have used the \cite{chang15} catalog to obtain infrared (IR)-based star formation rates (SFR). The catalog contains 858,365 galaxies within the SDSS spectroscopic sample as compiled in the New York University Value-added Galaxy Catalog \citep[NYU-VAGC;][]{blanton05, adc08, padmanabhan08} and cross-matched with the ALLWISE (Wide Field Infrared Survey Explorer) source catalog. Unlike optical emission line SFR estimates, \cite{chang15} utilized mid-IR data from full WISE photometry and employed an SED fitting technique to estimate stellar mass and  star formation rates. Their modeling is based on the MAGPHYS library \footnote{\href{http://www.iap.fr/magphys/}{http://www.iap.fr/magphys/}} (MAGPHYS contains 50,000 stellar population template spectra and 50,000 $\rm PAH+dust$ emission template spectra) and is applied to all z < 0.2 galaxies with good WISE photometry ($\rm FLAG\_W=1$ or 2), and good-quality SED fits ($\rm FLAG\_CHI2=1$). None of the objects in our sample features AGN-like WISE colors, hence it is unlikely that AGN contamination may be significantly biasing the SED-based SFR estimates. We have used the public \cite{chang15} catalogs \footnote{http://irfu.cea.fr/Pisp/yu-yen.chang/sw.html}. Details are given in \cite{chang15}.


\section{Method} \label{sec:method}

The identification of red geysers is based on the optical resolved spectroscopic data from MaNGA. Sub-section \S \ref{subsub:redgeys} describes the conditions and criteria that have been used to select our sample. Matched control sample galaxies have been selected from the full galaxy sample via the method discussed in \S \ref{subsub:control}. A third category of galaxies, which we call the ``H$\alpha$-disturbed'' class as described in detail in \S \ref{subsub:disturb}, consists of galaxies that are not classified as geysers due to the absence of the characteristic bi-symmetric pattern in resolved optical emission map, but they show kinematic and emission line properties that are quite different from typical quiescent galaxies. They have distinct regions of enhancements in H$\alpha$-equivalent width map, along with high spatially resolved gas velocity values in comparison to stellar velocity. The H$\alpha$-disturbed class forms a separate category, distinct from both the red geyser and control samples. We perform aperture photometry (described in detail in \S \ref{sub:photometry}) on the FIRST radio cutouts for all galaxies using an aperture size of 10$''$ diameter, to obtain radio flux values and the associated photometric errors. The galaxies which satisfy the condition $S/N > 3$ are classified as radio-detected with a confidence level of 3$\sigma$. Since the detection threshold of the VLA FIRST survey is shallow ($\sim$1 mJy), many galaxies might lie below the sensitivity limit. Section \S \ref{sub:photometry} describes the stacking analysis that allows us to constrain the average radio flux for samples of galaxies that are undetected individually. The median-stacked FIRST images for our sample provide greater signal-to-noise with typical rms $\approx$ 10 $\mu$Jy.

\subsection{Sample Selection} \label{sub:sample}

In this section we describe the identification of red geysers, the selection of matched control sample galaxies and characterization of the H$\alpha$-disturbed galaxies.

\subsubsection{Red geysers} \label{subsub:redgeys}

\begin{figure*}[h!!!] 
\centering
\graphicspath{{./plots/}}
\includegraphics[scale=.8]{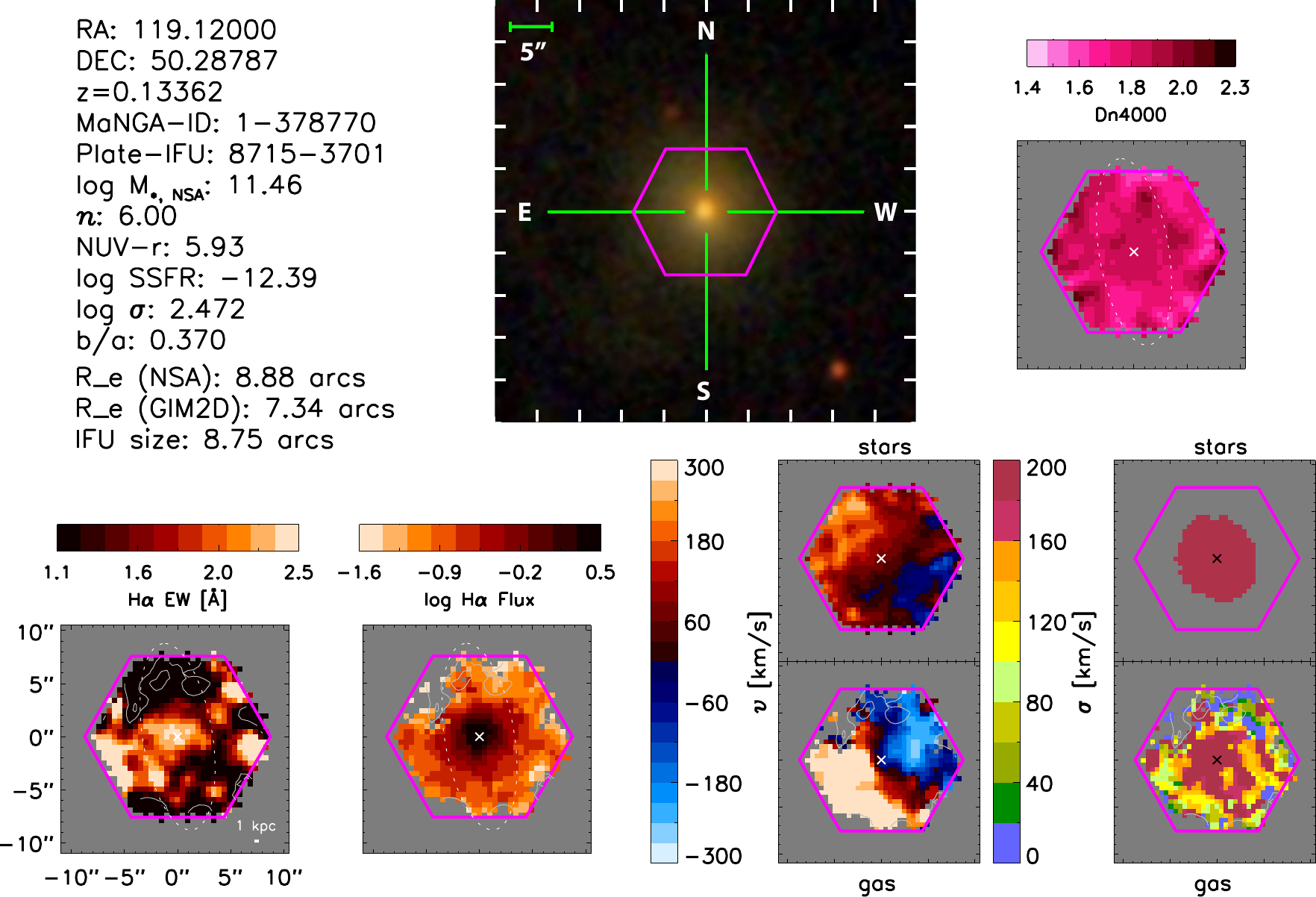}
\caption{ A typical red geyser included in our sample. The data has been obtained from MaNGA Integral Field spectroscopic observations. The panel on the center shows the optical image of the galaxy (MaNGA-ID: 1-634825). The magenta hexagon marked in the image is the extent of the MaNGA fiber bundle. In the other panels, as labelled, we have shown the H$\alpha$-flux map, Equivalent width map, Dn4000 absorption map, the velocity maps of gas and the stars along with their dispersion. As described in \S\ref{subsub:redgeys}, this galaxy satisfies all the conditions that we use to classify an object as red geyser. Specially notable is the bi-symmetric pattern in the equivalent width map of H$\alpha$ and the kinematic axis align perfectly with the gas velocity field. 
\label{fig:prototype}}
\end{figure*}

\begin{figure*}[h!] 
\centering
\graphicspath{{./plots/}}
\includegraphics[scale=.75]{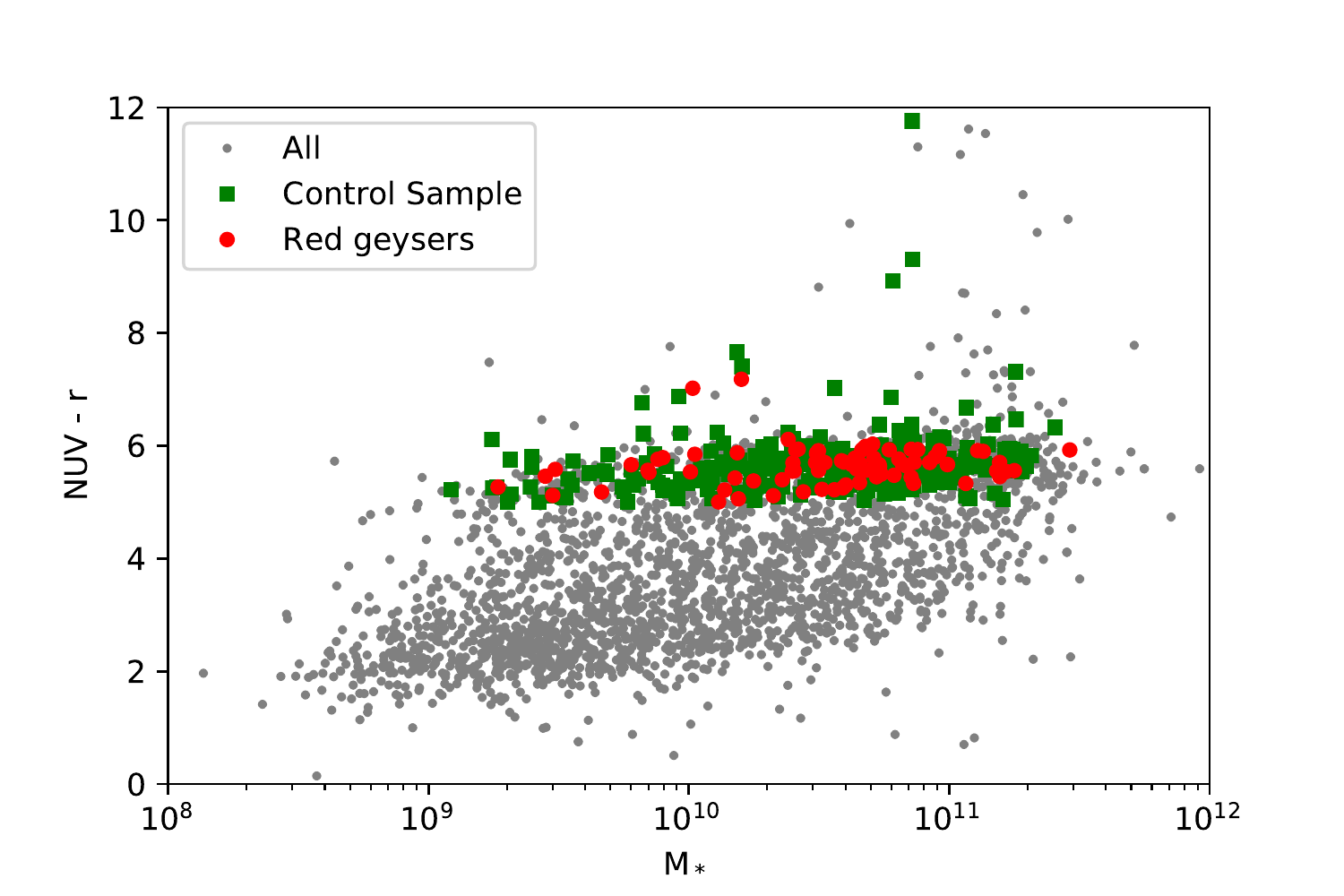}
\caption{The rest-frame $NUV-r$ color vs. stellar mass ($\log~M_*$) diagram of the MaNGA sample, with the red geysers in red circles and the control galaxies in green squares. Quiescent galaxies are clustered in the upper part of the $\rm NUV-r$ distribution; we define $\rm NUV-r>5$ as a conservative boundary of quiescent galaxies. Galaxies with $\rm NUV-r~>$~8 are undetected in the NUV data.
\label{fig:nuvrm}}
\end{figure*}

Red geysers are visually selected based on their characteristic features, as described in \cite{cheung16}. Red geysers are red galaxies defined by rest frame color $NUV-r>5$ (Fig \ref{fig:nuvrm}). The specified UV-optical color cut selects predominantly quiescent galaxies \citep{salim05, salim07, salim09}. In MPL-5, 40\% of all the targets fall in the quiescent category by the specific color cut. The SFR estimate derived from full SED fitting from optical to mid infrared data \citep{chang15} are small for the red geysers (96$\%$ of the sample has log SFR [$\rm M_\odot/yr] < -2$, shown later in Fig \ref{fig:wise_sfr}) to ensure quiescence and remove possible obscured star formation. Additionally, the spatially resolved map of EW of the Dn4000 absorption feature are high, roughly $>1.4$~\AA, thus ensuring the absence of young stars in the galaxy. The red geysers show narrow bi-symmetric patterns in the ionized gas emission as observed in the EW maps of strong emission lines like H$\alpha$ and [OIII]. These patterns line up approximately with the gaseous kinematic axis and show a misalignment with the stellar kinematic axis, but we pay close attention to cases where the mis-alignment of the stellar and gas velocity field is $0^{\circ}$, $180^{\circ}$ or $90^{\circ}$ in order to exclude embedded co-rotating, counter-rotating and polar gas disks. Another important defining property of the red geysers is that they have high spatially resolved gas velocities which can reach a maximum value of $\sim \rm 250-300~km~s^{-1}$, much higher compared to stellar velocities; as well as high gas velocity dispersion values (maximum $\sim \rm 200~km~s^{-1}$). Hence the observed second moment of the velocity  ($V_{\rm rms}\equiv \sqrt{V^2 + \sigma^2}$) of the ionized gas typically exceed the second velocity moments of the stars by $\rm 80-100~km/s$, suggesting that the ionized gas kinematics in these galaxies cannot be explained by gravitationally-bound orbits alone. For the prototype red geyser, \cite{cheung16} performed detailed Jeans Anisotropic Modeling (JAM) and used the dynamically constrained potential to get a rough estimate
of escape velocity, V$_{\rm esc} \sim 400\pm 50~\rm km/s$. They found that roughly 15-20\% of the
gas would exceed the escape velocity. A typical example of a red geyser is shown in Fig~\ref{fig:prototype}. Further details of the selection procedure will be described in Bundy at al. (in prep). 

Accreted gas disks in early type galaxies \citep[e.g.,][]{chen16} can sometimes produce similar gas velocity gradients like the red geysers, due to rotation of the gaseous material in the disk. A few edge-on disks show a bisymmetric pattern in EW map similar to the red geysers. Hence, we include a few steps in our visual identification, to exclude galaxies with a visible disk component or dust lanes apparent in the optical SDSS image. We discard edge-on galaxies with axis ratio $\rm b/a<0.3$. We also checked the galaxy specific stellar angular momentum ($\lambda_{Re}$) and ellipticity ($\epsilon$) from the extensive catalog in \cite{graham18}. Convincingly, we find that 95$\%$ of the red geysers are fast rotator early-type galaxies. Our control sample galaxies are of similar nature, 97$\%$ of which are fast rotators according to \cite{graham18}. Since the fast rotators have stellar disks and are axisymmetric, this implies that a gas disk cannot be in equilibrium if it is misaligned with the stellar kinematic position angle (PA), thus ruling out the possibility of disks to be the source of the observed misalignment in the red geysers.  Additionally, to avoid galaxies with rotating disk from being included in the red geyser sample, any galaxy showing a very low value of average gas velocity dispersion through out the galaxy ($< \rm 60~km~s^{-1}$ which is roughly the average dispersion value observed in polar disks), has been discarded from our red geyser sample completely, even if it shows other convincing features of a red geyser. 

As described in \cite{cheung16}, the gas velocity fields of the red geysers are poorly fit by flexible disk rotation models. The error-weighted average residual, characterizing the goodness-of-fit, place the red geysers among the worst 5\% of fitted MaNGA galaxies with disk-like kinematics. The resolved spectral line ratios land predominantly in the LINER region in the Baldwin, Phillips \& Telervich (BPT) diagram, predicting that the ionizing source is mostly post-asymptotic giant branch (AGB) stars and/or AGN.

To summarize, the red geysers in our chosen sample have the following characteristic features:
\begin{itemize}[noitemsep]
\item Quiescent nature with rest frame color $NUV-r>5$.
\item Bi-symmetric emission feature in H$\alpha$-EW resolved map.
\item Rough alignment of the bi-symmetric feature with the ionized gas kinematic axis, but misalignment with stellar kinematic axis.
\item High spatially resolved gas velocity values, typically reaching a maximum of $\pm$~300~km/s, which are greater than the stellar velocity values by atleast a few factors. 
\item Very low star formation rate with typical value of log SFR [$\rm M_\odot/yr] < -2$. 
\end{itemize}

Currently, our sample has 84 red geysers, which accounts for $\approx8\%$ of quiescent MaNGA galaxies (defined as $NUV-r>5$, see Section \S\ref{subsub:control}). 

\subsubsection{Control Sample} \label{subsub:control}

We create a control sample of quiescent galaxies with $ NUV-r>5$ (shown in Fig \ref{fig:nuvrm}), which are matched in global properties but do not show the resolved geyser-like features described in \S\ref{subsub:redgeys}. 

\begin{figure*}[h!!!] 
\centering
\graphicspath{{./plots/}}
\includegraphics[scale=.8]{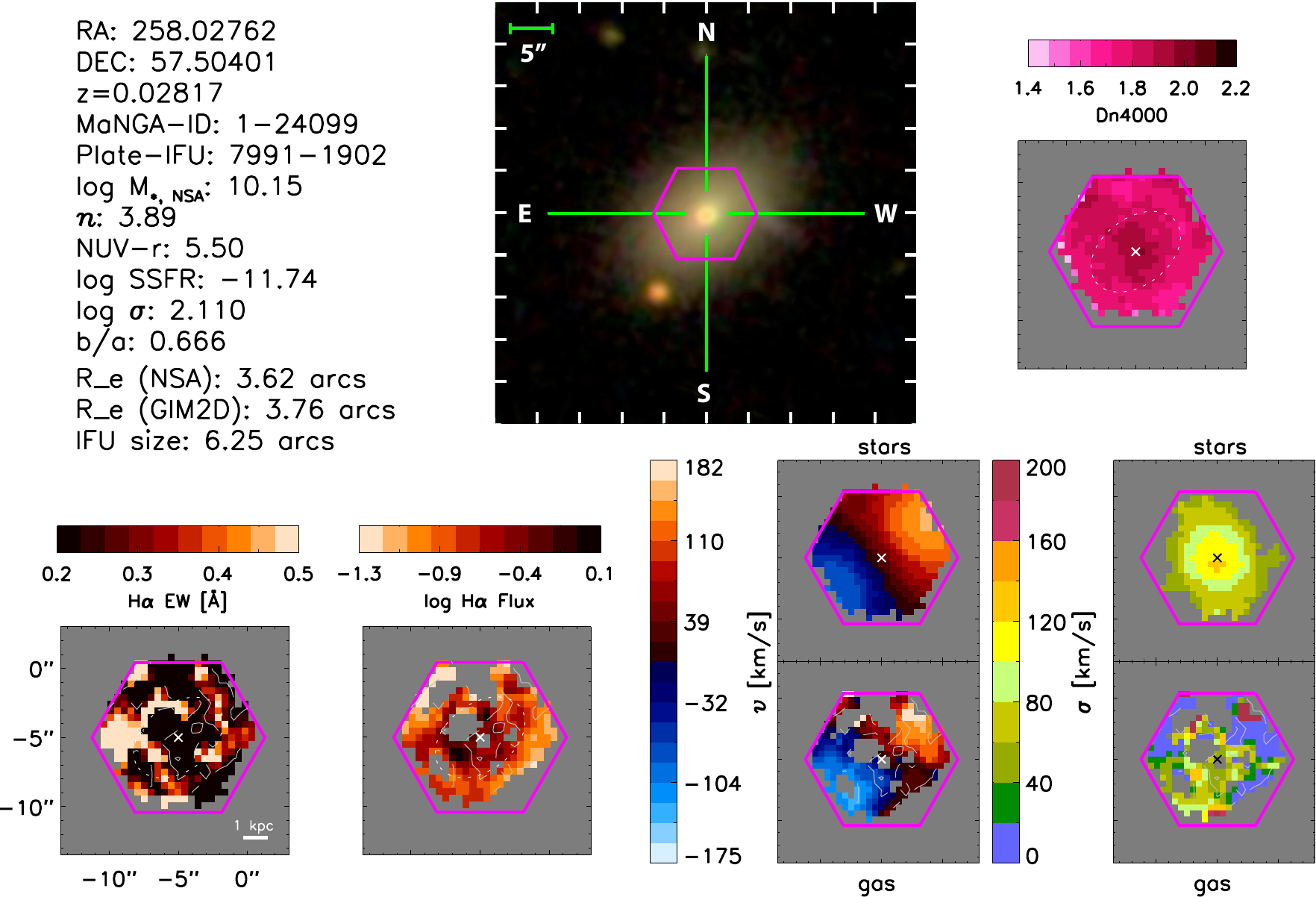}
\caption{ A typical control galaxy chosen in our sample. The data has been obtained from MaNGA Integral Field spectroscopic observations. The panel on the center shows the optical image of the galaxy (MaNGA-ID: 1-24099). The magenta hexagon marked in the image is the extent of the MaNGA fiber bundle. In the other panels, as labelled, we have shown the H$\alpha$-flux map, Equivalent width map, Dn4000 absorption map, the velocity maps of gas and the stars along with their dispersion. As described in \S\ref{subsub:control}, this galaxy is red with NUV-$r>\rm 5$, has a very low value of star formation and it is relatively face-on with $\rm b/a>0.3$. This galaxy is clearly not a red geyser as it does not satisfy any of the red geyser features described in \S\ref{subsub:redgeys}, so it can safely be included in the control sample.
\label{fig:control}}
\end{figure*}

For each red geyser, we match up to five unique quiescent galaxies with the following criteria:

\begin{itemize}[noitemsep]
\item $\lvert \log ~M_{*, \rm ~red~geyser}/M_{*, \rm ~control} \rvert < 0.2$ dex
\item $\lvert z_{\rm red~geyser} - z_{\rm control} \rvert < 0.01$
\item $\lvert b/a_{\rm red~geyser} - b/a_{\rm control} \rvert < 0.1$,
\end{itemize}

\noindent where $M_*$ is the stellar mass, $z$ is the spectroscopic redshift, and $b/a$ is the axis ratio from the NSA catalog. Stellar mass and redshift have been shown to correlate with radio emission and thus must be controlled for \citep[e.g.,][]{condon84, dunlop90, best05}. We also control for axis ratio so that we do not compare potentially dust-reddened edge-on galaxies with the relatively face-on red geyser galaxies. This matching technique results in 260 unique control galaxies. Fig~\ref{fig:control} shows an example of a typical quiescent galaxy from the control sample. 

\begin{figure*}[h!!!] 
\centering
\graphicspath{{./plots/}}
\includegraphics[scale=.75]{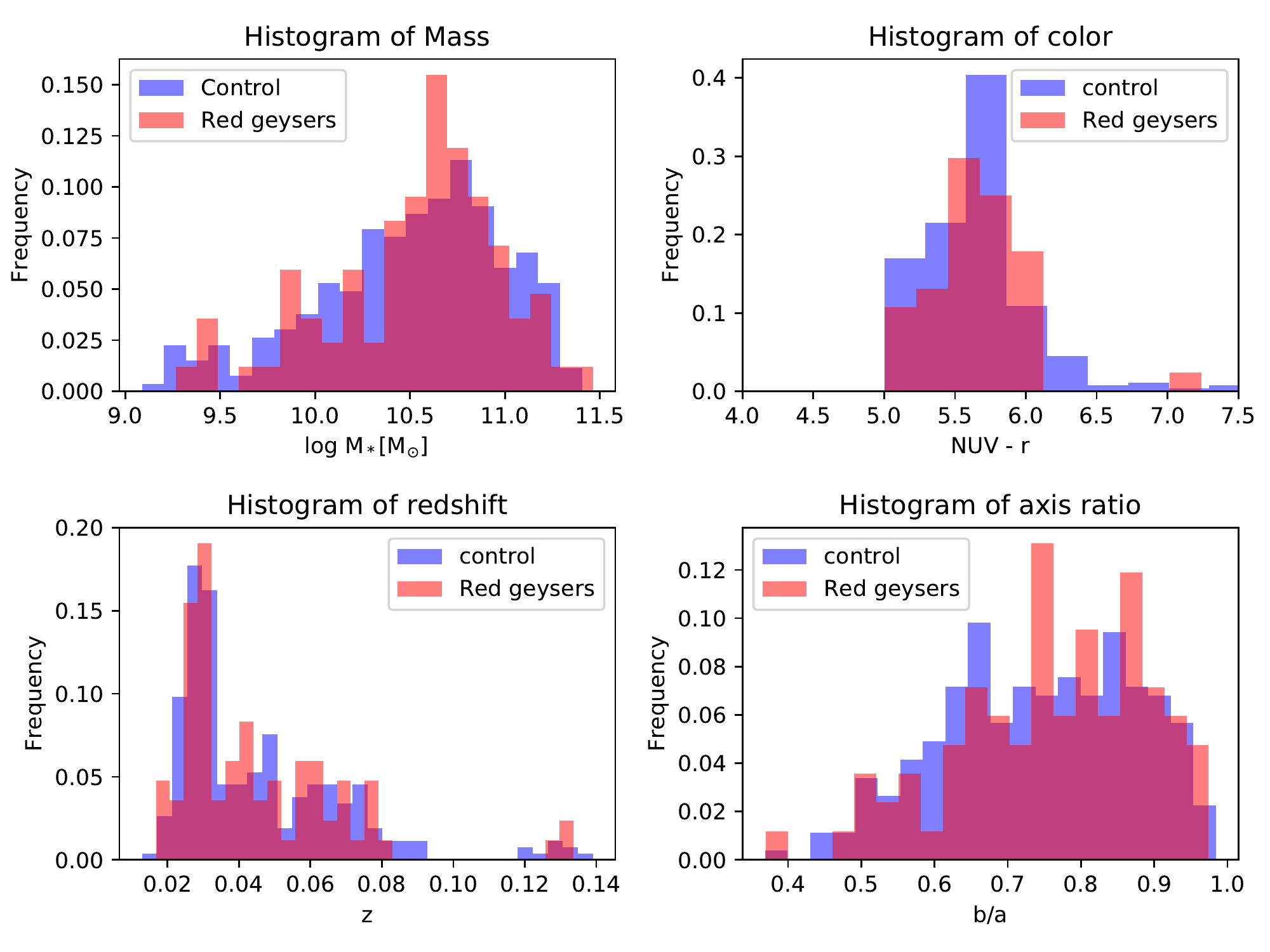}
\caption{ Comparison of global properties of red geysers with our chosen control sample. Normalized histograms of the red geysers and control galaxies in: stellar mass ($\log~M_*$), rest-frame $NUV-r$ color, redshift ($z$), and axis ratio ($b/a$). The red geyser sample distribution are shown in red, while the control sample properties are shown in blue. We see similar distribution for red geysers and control sample properties, as expected from our method of selection of control sample. 
\label{fig:all}}
\end{figure*} 

Fig.~\ref{fig:all} compares the global galaxy properties of the control sample and the red geysers.  The red geysers (red) and the control sample (blue) are well-matched in all four parameters-- stellar mass, redshift, color and axis ratio as expected.

\subsubsection{H$\alpha$-Disturbed Galaxies} \label{subsub:disturb}

\begin{figure*}[h!!!] 
\centering
\graphicspath{{./plots/}}
\includegraphics[scale=.8]{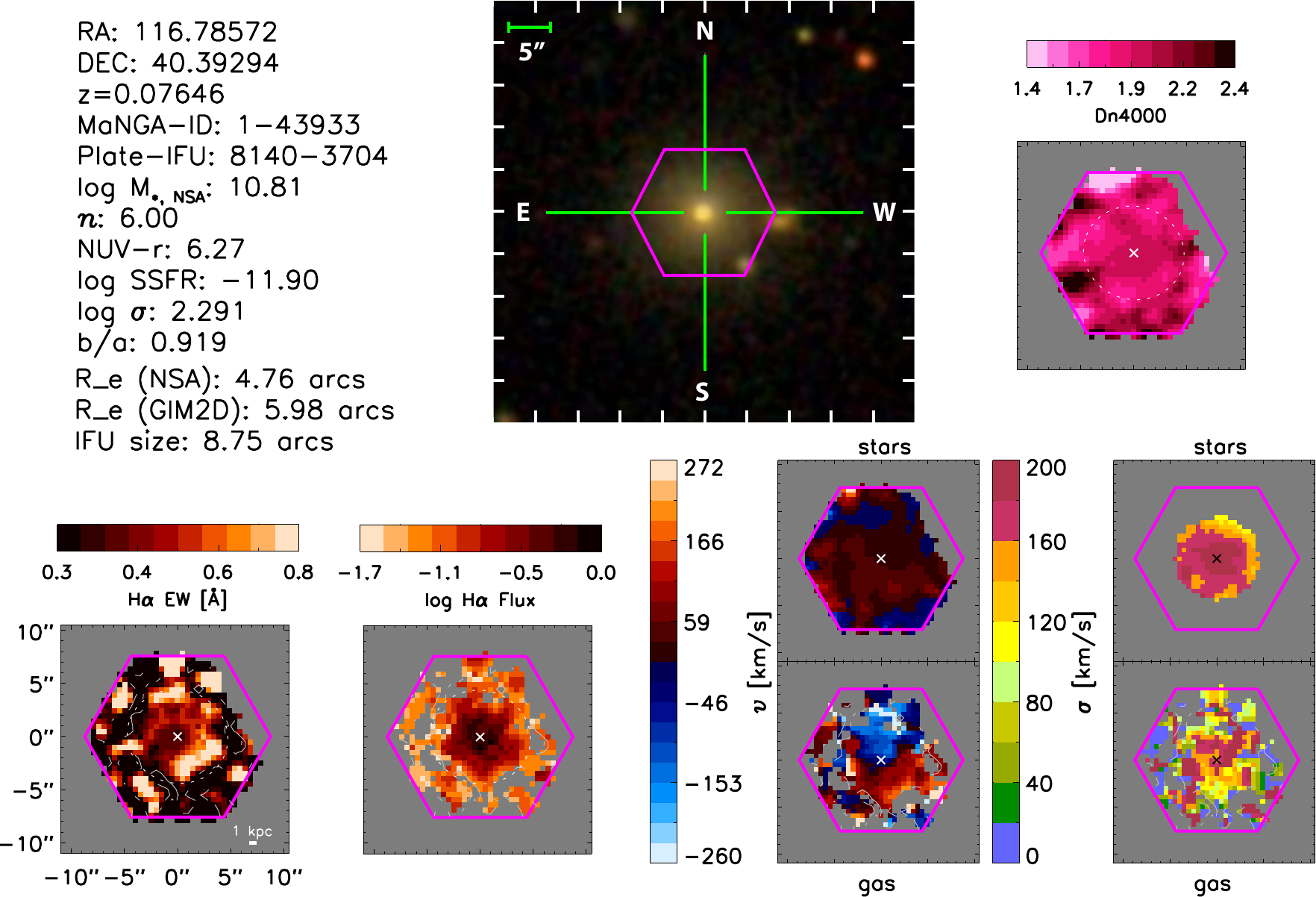}
\caption{ A typical disturbed galaxy as described in \S\ref{subsub:disturb}. The data has been obtained from MaNGA Integral Field spectroscopic observations. The panel on the center shows the optical image of the galaxy (MaNGA-ID: 1-43933). The magenta hexagon marked in the image is the extent of the MaNGA fiber bundle. In the other panels, as labelled, we have shown the H$\alpha$-flux map, Equivalent width map, Dn4000 absorption map, the velocity maps of gas and the stars along with their dispersion. As described in \S\ref{subsub:disturb}, this galaxy cannot be called a promising geyser candidate because of the lack of the signature bisymmetric pattern, but the kinematics indicate a difference from ordinary control sample. It has been classified as a third ``H$\alpha$-disturbed'' category to separate from the geyser and the control sample population
\label{fig:dist}}
\end{figure*}

During the course of visual inspection, we have discovered another category of galaxies which we will hereby refer to as ``H$\alpha-$disturbed''. Fig~\ref{fig:dist} shows an example. The gas content of these galaxies is comparable to the red geysers (median H$\alpha$ EW value $\rm>0.5~$\AA~similar to $\sim\rm 0.8~$\AA~in the red geysers) but the H$\alpha$ equivalent width maps do not show the clear bisymmetric patterns of a red geyser. They show twisted, disturbed H$\alpha$ EW maps, sometimes with individual blobs of gas that are found throughout the galaxy. 90\% of the sample have spatially resolved gas velocity values reaching a maximum of $\sim \rm 250~km/s$, which are high compared to the stellar velocities which lie within $\pm$~60~km/s. Some of them have high gas velocity dispersion, upwards of $\sim\rm200~km~s^{-1}$ as seen in red geyser population. 
We found 60 such H$\alpha-$disturbed candidates from $\sim~900$ MaNGA quiescent population, and we treat them as a separate third category different from both the red geyser and control samples.

\subsection{FIRST Radio Photometry and Stacking} \label{sub:photometry}

To obtain the radio flux, we perform aperture photometry on the FIRST cutouts for 78 out of 84 red geysers, 260 control galaxies and 57 out of 60 H$\alpha$-disturbed galaxies which have FIRST coverage. We first determine which FIRST tile (of dimension $34.5' \times 46.5'$) a specific galaxy falls on. If a galaxy is located too close to the FIRST tile edge (less than 10$''$), that galaxy is discarded. We extract a small cutout 50$\times$50 pixels wide (each pixel is 1.8$''$) centered on the galaxy of interest. We use a circular aperture of 10$''$ diameter centered on the galaxy and sum the radio flux values within. For our target galaxies which are located roughly at redshift $\sim0.03$, 10$''$ aperture corresponds to 6 Kiloparsec (Kpc) on the sky and hence is a reasonable choice as aperture size. We have defined the criteria for radio detection to be $S/N > 3$.  
We then perform a median stack of the FIRST images associated with the three samples described in \S \ref{sub:sample}. To ensure that our results are not biased by a few radio bright sources, we have made separate stacks of radio flux with the individually radio detected sources removed.

We have also tested that our stacked radio signal is not an artifact of faulty FIRST tiles by median stacking random cutouts within a radius of 75$''$ in the same FIRST tile where the galaxy is located. We would expect these ``blank'' stacks to have pure white noise with no radio signal. 

\begin{figure*}[h!!!] 
\centering
\graphicspath{{./plots/}}
\includegraphics[scale=.5]{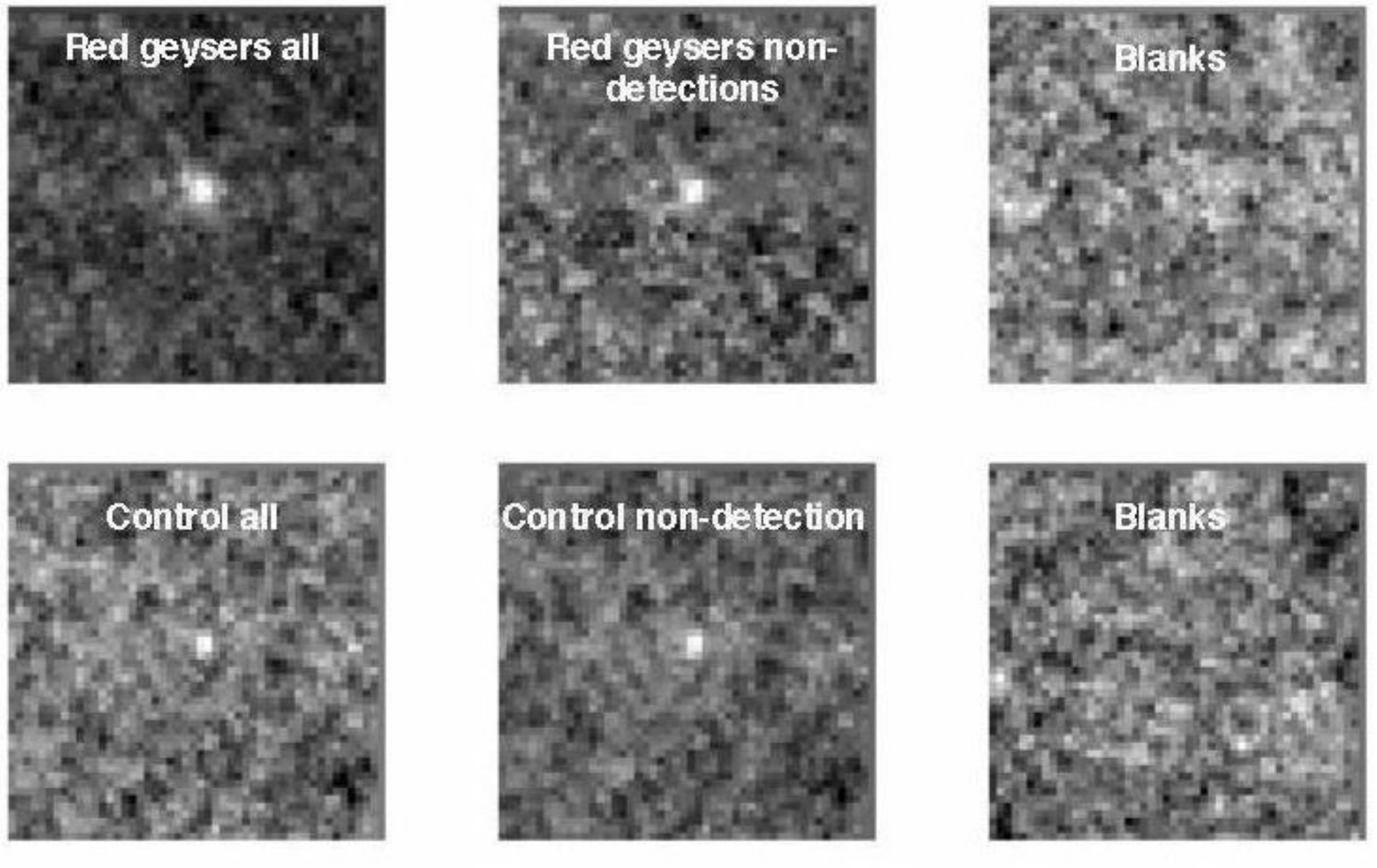}
\caption{ The median stacked images of red geysers (top) and control sample (bottom). The middle panels show the non-radio detected stacked images for the red geysers (middle) and the control (bottom middle), where all radio detected sources have been excluded. The blank stacks are shown in (top right) and (bottom right) panels. 
\label{fig:stack}}
\end{figure*}

Fig.~\ref{fig:stack} shows the images of the median stacked flux of these four samples --- (1) the red geysers, (2) the control sample, (3) the non-radio-detected red geysers, and (4) the non-radio-detected control sample. The rightmost panel in both the rows show the blank stacks. Reassuringly, the blank stacks show no signal.

We perform additional separate stacks controlling for ionized gas content and star formation rates in the control galaxies to see their effect on the radio output. Details of our findings are given in \S \ref{sec:results}.

In order to account for the photometric error as well as the systematic error due to sample construction, we perform a bootstrap analysis on all our samples. We construct 1000 random samples with replacement with the same size as each sample and compute the stacked radio flux as before. We take the standard deviation of the resulting flux distributions ($\sigma$) to be the estimate of the error on the stacked flux measurements.

\section{Results} \label{sec:results}

\subsection{Radio detection of red geysers vs. control sample}

 We have crossmatched FIRST radio detections with our sample of red geysers and control galaxies. 12 $\pm$ 3 out of 78 red geysers ($\sim~15 \%\pm 4 \%$) are found to be radio-detected, where quoted errors are obtained from standard Poisson statistics. Among the control sample, 14 $\pm$ 4 out of 260 are radio detected with a detection fraction $\sim 5\% \pm 1.5\%$.  Red geysers show a 3 times higher radio detection rate compared to our control sample with a significance level of 5$\sigma$. We also find that the radio detected red geysers make up an appreciable fraction ($\sim10\%$) of the red MaNGA galaxies which are  radio-detected by FIRST survey. This fraction increases to $\sim20\%$ when the H$\alpha$-disturbed category galaxies are included along with the red geyser population. If we limit our sample to $\rm log~(M_\star/M_\odot) < 11$, the detection rate of red geysers and H$\alpha$-disturbed goes up to $\rm 40\%$. 
 

\subsection{Stacked radio activity of red geysers vs. the control sample} \label{sub:radio_comparison}

\begin{figure*}[h!!!!!!!!!!!!!!!] 
\centering 
\graphicspath{{./plots/}}
\includegraphics[scale=.85]{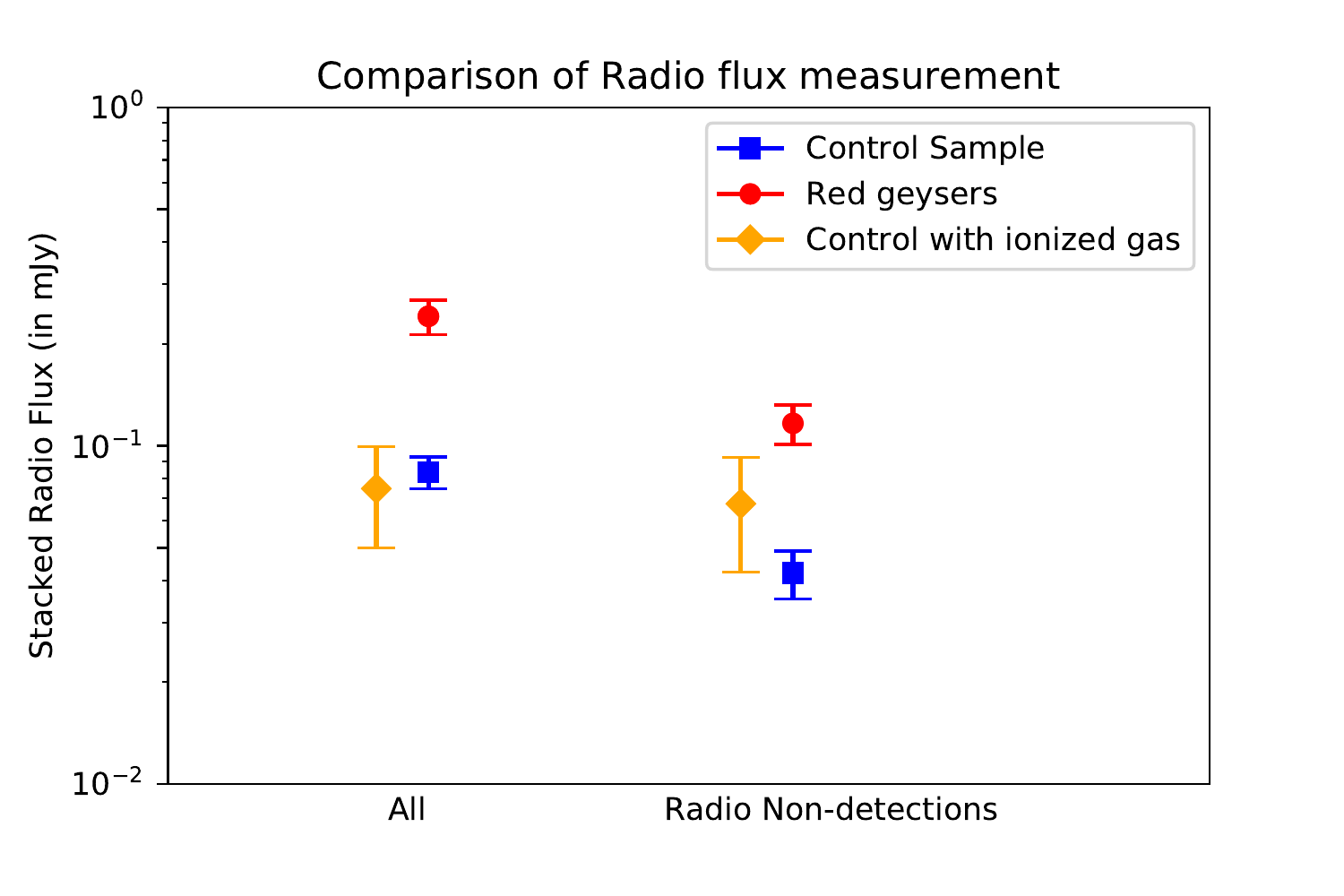}
\caption{ The median stacked radio flux obtained from the stacked sample of red geysers (shown in red circles) and control sample (blue squares). 
{\it ``All''} represents the stacks where the entire sample has been included for both red geysers and control, while {\it ``Radio Non-detections''} indicate the stacks where the individually radio-detected sources have been removed. The condition of radio detection of a source has been defined as $\rm SN>3$. ``Control with gas"-- marked in yellow diamonds-- shows a specific subset of control galaxies when we additionally controlled for ionized gas (described in \S\ref{sec:results} in details). 
The red geyser sample show an enhanced radio flux compared to control sample galaxies and the 
presence of higher amount of  ionized gas in the control sample does not necessarily affect the radio-detection rate. The spaxel by spaxel equivalent width information have been obtained from the MaNGA DAP (Data analysis Pipeline) and they have been averaged over the spatial extent of 1.5 effective radii to obtain the mean EW value for a particular galaxy.
\label{fig:radio}}
\end{figure*}

Fig~\ref{fig:radio} shows the first main result of our analysis. We compare the median-stacked radio fluxes of the red geysers (red circles) with that of the control sample (blue squares). 
Data points in the column marked ``All'' indicate the median fluxes when the entire sample of geysers and control samples are included in the stack. In the column labelled ``Radio non-detection'' we have excluded radio bright red geysers and control galaxies. We see that for both these cases, the red geyser radio fluxes, obtained from median stacking, are 3 times higher than the control sample at greater than $99.99\%$ confidence ($>5\sigma$).

We additionally control for the presence of ionized gas in our sample. We obtain H$\alpha$ equivalent width (EW) measurements from the MaNGA DAP. The mean value obtained by averaging the EW (H$\alpha$) values of all spaxels in a particular galaxy within 1.5 effective radii is used as the mean EW value, and a proxy for ionized gas content. The control galaxies show an average value of 0.3~\AA, somewhat lower than the corresponding 0.8~\AA\  seen in the red geyser sample. To compare against galaxies with similar equivalent width values, we select an additional control sample with EW > 0.5~\AA\ (stacks marked with yellow diamond points). We see that even the radio stack of control galaxies having a comparable level of ionized gas, has a value about 3 times less than that of the red geyser stack. In addition to that, the stacked radio flux for the control galaxies with ionized gas does not show much difference for ``All'' and ``Radio Non-detections'' sample, which implies that presence of higher amount of  ionized gas in the control sample does not necessarily affect the radio-detection rate.

 The detailed implications of these findings are summarized in \S \ref{sec:conclusion}.

\subsection{Dusty star-formation} \label{sub:dust}

As described in \S\ref{subsub:redgeys}, we set a color cut of rest frame $NUV-r>5$, and exclude edge on galaxies with $\rm b/a<0.3$ to avoid possible radio contamination by star formation to the radio flux. However, UV wavelengths are susceptible to dust attenuation and may not reveal heavily obscured star formation \citep[e.g.,][]{calzetti00}. Here we use the SDSS+WISE \cite{chang15} catalog for obtaining star formation rates (SFR) based on IR fluxes that are sensitive to dusty star formation. \cite{chang15} has utilized the full WISE photometry to model the spectral energy distribution (SED) in optical through mid-IR bands and obtained updated measures of mass and SFR.  

\begin{figure*}[h!!!!!!!!!!!!!!!] 
\centering
\graphicspath{{./plots/}}
\includegraphics[scale=.9]{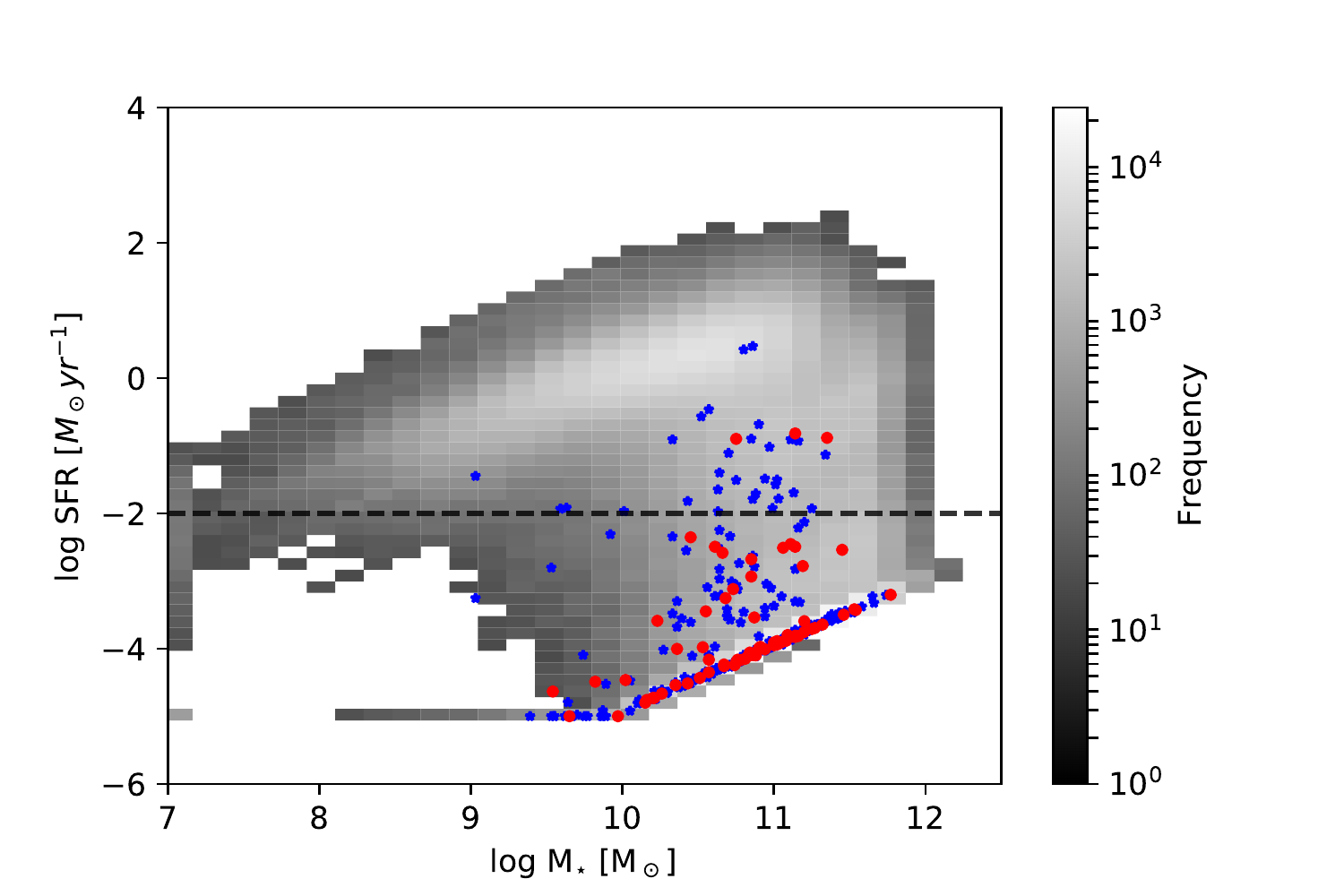}
\caption{ The figure shows the log SFR vs log M$_\star$ as obtained from SDSS+WISE catalog of \cite{chang15}. The gray 2D histogram shows all the galaxies in the catalog with $\rm0.01<z<0.1$. The red circles and blue stars signifies red geyser and control sample galaxies respectively. Most of the galaxies in our chosen sample have a low log SFR value, $< -2$ M$\odot$/yr. 
\label{fig:wise_sfr}}
\end{figure*} 


Fig ~\ref{fig:wise_sfr} 
show the log SFR vs log M$_{\star}$ 
diagram of the galaxies from the \cite{chang15} catalog. We see that the majority of red geyser and control galaxies lie in the non-star forming region, with low values of SFR. 
To ensure that our result is not affected by radio contamination from dusty star formation, we have redone the stacking analysis after excluding galaxies that have log SFR~[$\rm M_\odot/yr]> -2$. This cut removes 3 red geysers and 30 control sample galaxies. Fig ~\ref{fig:radio2} shows the median stacked radio flux in the column labelled ``Non-Starforming". We conclude that our results are not affected by contamination from dusty star formation. 

WISE colors can be used to detect strong nuclear heating associated with bright AGNs or quasars at the center of the host galaxy. According to \cite{yan13}, $\rm W1~(3.4 \mu m) - W2~(4.6 \mu m) > 0.8$ presents an efficient mid-IR color based selection criteria for luminous AGN and quasars. Most of the red geysers and control sample have $\rm 0.6 < W1 - W2 < 0.7$ with very few (1 or 2) having a value $>$~0.8. This lends confidence to the ability of the WISE data to constrain obscured star formation in these galaxies, as there is no AGN contamination present in the SEDs of these galaxies. We trust the SED-based SFRs from \cite{chang15}.

\subsection{Stacked flux of H$\alpha-$Disturbed category} 

\begin{figure*}[h!!!!!!!!!!!!!!!] 
\centering
\graphicspath{{./plots/}}
\includegraphics[scale=.85]{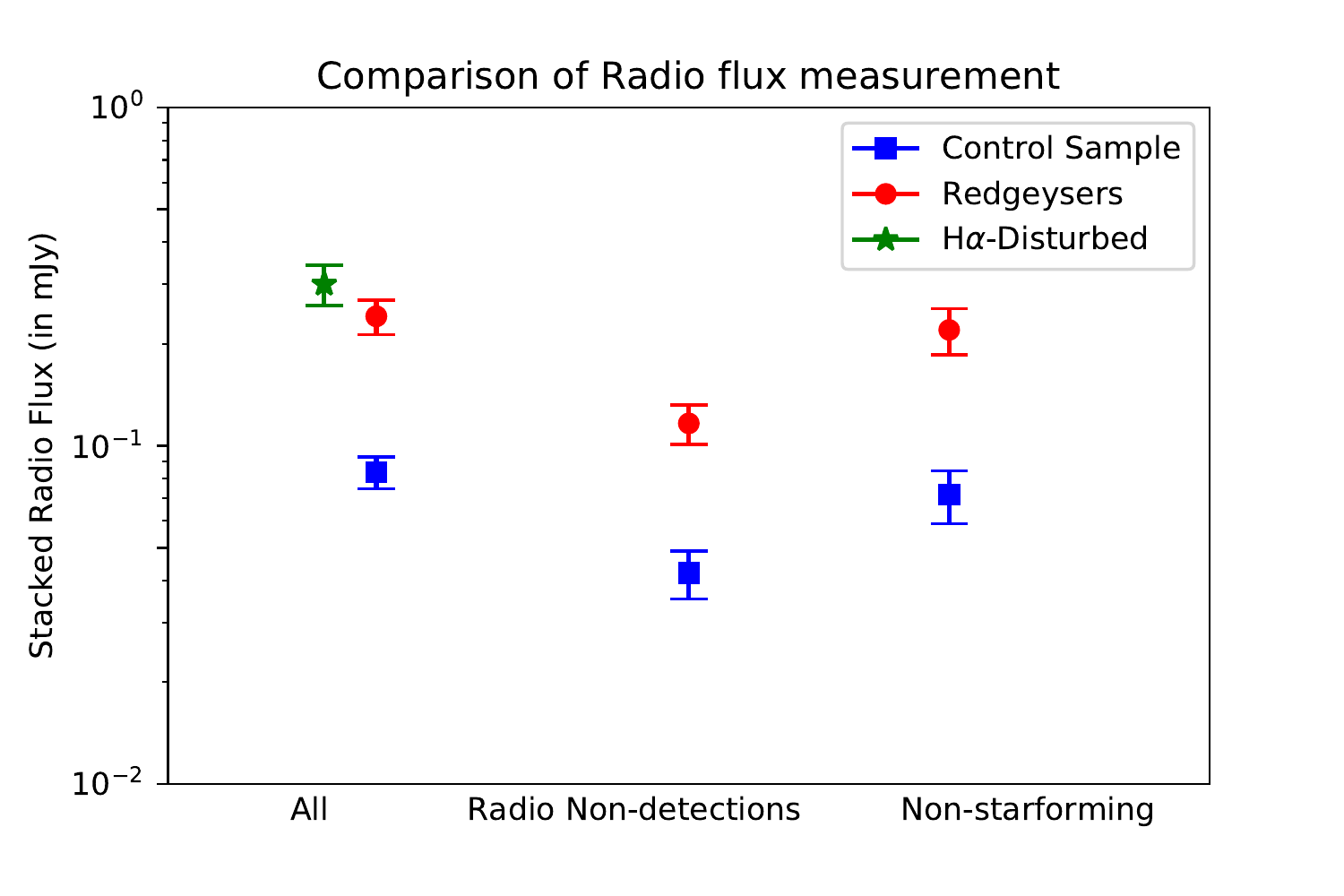}
\caption{ The median stacked radio flux obtained from the stacked sample of red geysers (shown in red), H$\alpha$-disturbed (pink) and control sample (blue). 
The leftmost panel shows the stacks for the entire sample of red geysers (shown in red circles), control sample (shown in blue squares) and the H$\alpha$-disturbed category (shown in green star). {\it ``Radio Non-detections''} panel shows the stacked radio flux for the geysers and the control sample where the individually radio-bright ones, satisfying the criteria $\rm S/N >3$, have been removed. The red geyser and the control sample have been cross-matched with SDSS+WISE catalog of \cite{chang15}. Galaxies with $\rm log~SFR > -2 M\odot/yr$ have been removed and re-stacked. They constitute the  {\it ``Non-Starforming''} category shown in the rightmost panel of the plot. In all the cases, the median stacked radio flux is higher for the red geyser sample compared to the control sample by $>5\sigma$. 
\label{fig:radio2}}
\end{figure*}

In Fig~\ref{fig:radio2}, the stacked flux for the galaxies in the disturbed category is shown in green star symbol. Remarkably these galaxies show a slightly higher value of median stacked radio flux than the red geysers. The disturbed EW maps and high gas velocity dispersions revealed by MaNGA data correlate with enhanced radio flux. We will discuss the implication of this finding in Section \S \ref{sec:conclusion}.

\section{Discussions \& Conclusion} \label{sec:conclusion}

We have performed a radio stacking analysis of 78 red geysers selected from the MaNGA survey which have FIRST coverage and have compared their median radio flux with a sample of quiescent galaxies matched in global integrated properties but are not classified as red geysers. The red geyser galaxies show significantly higher radio fluxes than the control galaxies, despite the fact that the red geyser selection is based on optical data alone.  This suggests a physical link between the optical features that identify red geysers and the presence of enhanced AGN activity, lending further support to the argument that AGN-driven winds are responsible. 

We have made several subdivisions  based on different physical criteria, to check our results:
\begin{itemize}
 \item We have performed the stacking for all galaxies both in the red geyser and control samples.
 
 \item We have performed the stacking for samples in which the radio detected sources are removed so that a few bright sources do not dominate the median stacked radio flux value.
 
 \item We have performed the stacking for galaxies with similar levels of ionized gas by imposing a cut on EW (H$\alpha$) value.
 
 \item We have performed the stacking for samples that exclude galaxies which show a high value of star formation from SDSS+WISE.
 
 \end{itemize}
 
 In all cases red geysers exhibit elevated radio flux values.
  
  Given our conservative $NUV-r$ color cut, the use of WISE mid-IR data (\S \ref {sub:dust}) and absence of star forming HII regions from resolved BPT diagrams, we can rule out star formation as the explanation for this enhanced radio flux. The other most likely sources are AGN activity or Supernova remnants. SN Ia remnants can induce radio synchrotron emission from shock-accelerated cosmic rays. However in our case, they are unlikely to be responsible for the increased radio signal in red geyser sample because our selection criteria do not involve any factors that may enhance or suppress the SN Ia rate. We have controlled primarily for the $M_*$, rest-frame $NUV-r$ color and age of the galaxy respectively. Thus there should be no difference in the frequency of SN Ia remnants between the red geysers and the control sample.
  
  \begin{figure*}[h!!!!!!!!!!!!!!!] 
\centering
\graphicspath{{./plots/}}
\includegraphics[scale=.85]{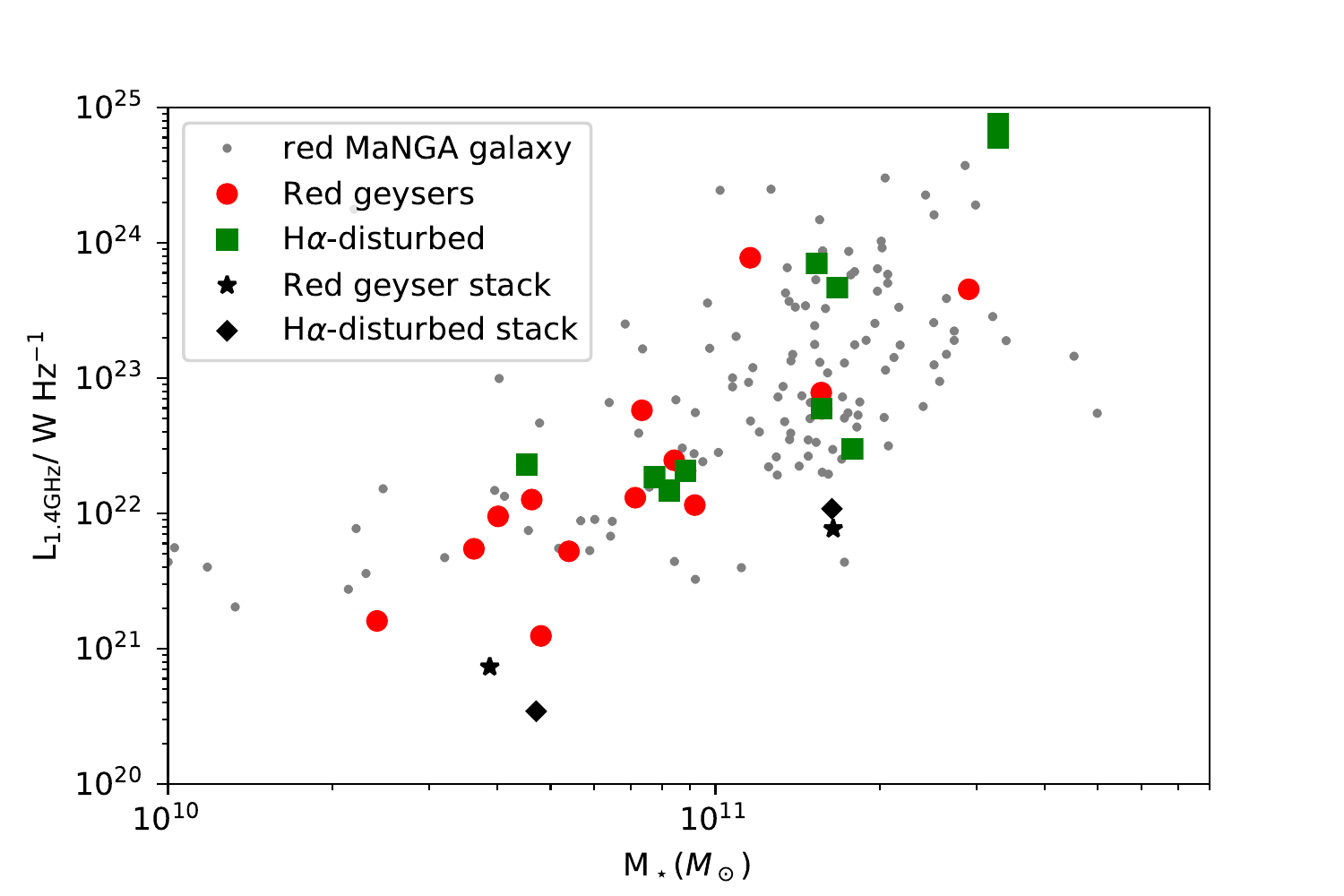}
\caption{ The figure shows the 1.4GHz radio luminosity versus stellar mass of radio detected red geysers (shown in red big circles), H$\alpha$-disturbed (shown in green squares) and the red ($\rm NUV-r > 5$) MaNGA galaxies (in gray small circles). This plot shows that the radio AGN population occupies two distinct regions in the luminosity stellar-mass space depending on the types of host quiescent galaxies. The lower mass regime ($\rm log~M_\star < 11$) is occupied by quiescent galaxies with optical emission line features (red geysers and H$\alpha$-disturbed) while in the higher mass region, we mainly find galaxies without detectable emission line features (similar to our control sample). The black stars and diamonds show the stacked radio luminosities from the entire sample (which includes both radio-detected and non-detected ones) of red geyser and H$\alpha$-disturbed galaxies respectively, in two mass bins.
\label{fig:Lradio}}
\end{figure*}

 We conclude that the enhanced radio emission of red geysers is due to the presence of radio-mode AGNs. To confirm this statement, we compare the expected SFR from the average radio luminosity from the entire red geysers sample with the observed SFR derived from full SED fitting of optical mid-infrared data from \cite{chang15} catalog. The mean radio luminosity density (L$_{\rm1.4GHz}$) obtained from the stacked integrated flux density is L$_{\rm1.4GHz} \sim \rm 2 \times 10^{21}~W~Hz^{-1}$ (obtained by averaging the values of radio luminosity of the red geysers in two mass bins shown by two black filled-stars in Fig \ref{fig:Lradio}). From the best-fit relation between 1.4 GHz radio continuum luminosity and the Balmer decrement corrected H$\alpha$ \citep{brown17}, we obtain a corresponding H$\alpha$ luminosity $\sim \rm 1.3 \times 10^{41}~erg~s^{-1}$. Using the known relation between SFR and H$\alpha$ luminosity \citep{kennicutt09, brown17} assuming a Kroupa intial mass function (IMF) \citep{kroupa}, we obtain an expected SFR from this radio emission $\sim \rm 1 M_{\odot}~yr^{-1}$. However, the observed mean SFR from the sample, as shown in Fig \ref{fig:wise_sfr}, is $\sim \rm 10^{-3}M_{\odot}~yr^{-1}$, which is much lower than expected, thus confirming AGN to be the primary source of radio emission rather than star formation in the red geysers. The AGN feedback can induce radio emission through their radio jets \citep{zensus97, falcke99}, their advection-dominated accretion flows (ADAFs; \citealt{narayan95, narayan00}), and/or their winds \citep{jiang10}.
 
 It stands to reason that the AGNs in the red geysers may act as the central powerhouse driving the ionized gas winds that signal the red geyser phenomenon.

It is interesting to consider how the H$\alpha$-disturbed galaxies fit in this context. These galaxies show a comparable (within uncertainty) or a slightly higher value of stacked radio flux compared to the red geysers. All of them show significant gas blobs in the H$\alpha$ EW maps. Some of them can be potential geyser candidates or relics from mergers or tidal interactions with other galaxies. The complex gas morphology might be a product of a multi-phase and clumpy interstellar medium, ionized by the central AGN. These blobs may form out of the geyser wind material after the central engine shuts down. They may also result from a less stable accreting source. Given the uncollimated and chaotic distribution of ionized gas, it seems unlikely that cool inflowing of material from a galactic encounter is responsible. There is also no indication that that H$\alpha$-disturbed galaxies have recently undergone a merger or interaction. Clearly more work is needed to understand them. 


 We would also like to highlight the handful of control galaxies with clear radio detections that are not classified as red geysers or as H$\alpha$-disturbed. These galaxies likely host a central active nucleus and exhibit significant emission line flux. They may mean any of the following:

\begin{itemize}
\item Our red geyser sample based on visual inspection is not a complete sample of AGN-driven ionized winds. Red geysers may be a special type of AGN wind phenomena. 

\item The AGNs in the control sample are too weak too drive out sufficient gas for detection at large radii.  

\item A time lag may exist between AGN triggering and the development of a large-scale wind. Those AGN hosted control galaxies may not be in the red geyser phase at the current epoch, but may have passed through this phase in the past, or might in the future. 
\end{itemize}

Fig \ref{fig:Lradio} shows the variation of radio luminosity (L$_{\rm 1.4GHz}$) with stellar mass (M$_\star$) for all the FIRST radio-detected quiescent galaxies in MaNGA sample. We see that radio-AGN population occupies two distinct regions in the plot depending on the properties of host quiescent galaxies. Radio-AGN in the galaxies showing optical emission line features (red geysers and H$\alpha$-disturbed) are found mostly at $\rm log~(M_\star/M_\odot) < 11$ (low mass end of the typical quiescent galaxy population) while the radio detection rate overall is higher for $\rm log~(M_\star/M_\odot) > 11$ by almost a factor of 1.4, compared to the lower mass galaxies.  One possibility is that red geysers and ``radio galaxies'' represent different AGN populations with different associated accretion histories and fueling mechanisms.  Alternatively, the declining presence of wide-scale ionized gas at higher stellar mass \citep{belfiore17} may simply hide the existence of AGN-driven winds at higher masses.

We can gain further insight by considering the average luminosities from our stacked samples in two stellar mass bins. Using the median redshift in each mass bin, we overplot the average luminosity of red geysers and H$\alpha$-disturbed galaxies on Fig~\ref{fig:Lradio} (shown in black filled-stars and diamonds respectively). The average luminosity has been obtained from the stacked radio flux that includes both radio-detected and non-detected sources. While radio-detected sources show a strong mass dependence, the average radio luminosity associated with red geysers show a slightly weaker dependence with stellar mass.  This suggests that a different kind of accretion physics may be at play.

Considering the two red geyser mass bins in Fig~\ref{fig:Lradio}, we see that the typical radio power of the red geysers is $\sim~10^{21}\rm\ W \ Hz^{-1}$(shown in Fig \ref{fig:Lradio} by the two black filled stars). From the best-fit linear relation between jet mechanical energy and the radio power from \cite{heckman14}, we get an estimate of the jet kinetic energy to be 3 $\times 10^{41}$ erg/s. The AGNs in the red geysers are low-luminosity sources and their mechanical
energy will be confined predominantly to size-scales of the host galaxy halo. Also, according to our interpretation, the short-lived geyser phase possibly occur in any red-sequence galaxy, hence the ``duty cycle'' represents the number of galaxies with an active red geyser phase in the present. Hence, if we assume that the observed fraction of red geysers represents their duty cycle, then these phenomena are present ~10\% of the time (higher, if we consider $\rm log~(M_\star/M_\odot) < 11$).  Multiplying this duty cycle by the typical jet kinetic power yields $\sim~$3$\times$10$^{40}$erg/s, an estimate of the AGN power averaged over long time scales.  We can compare this to the cooling rate implied from the X-ray gas in this stellar mass range \citep{best06,sullivan01}, which is similar, $\sim~$10$^{40}$erg/s.  This similarity provides further evidence that red geysers may play an energetically interesting role in the suppression of gas cooling and star formation at late times.\\
\\


NR thanks Professor Puragra Guhathakurta  for helpful comments and discussions. RAR acknowledges CNPq and FAPERGS.\\
Funding for the Sloan Digital Sky Survey IV has been provided by the Alfred P. Sloan Foundation, the U.S. Department of Energy Office of Science, and the Participating Institutions. SDSS-IV acknowledges
support and resources from the Center for High-Performance Computing at
the University of Utah. The SDSS web site is \href{http://www.sdss.org}{www.sdss.org}.

SDSS-IV is managed by the Astrophysical Research Consortium for the 
Participating Institutions of the SDSS Collaboration including the 
Brazilian Participation Group, the Carnegie Institution for Science, 
Carnegie Mellon University, the Chilean Participation Group, the French Participation Group, Harvard-Smithsonian Center for Astrophysics, 
Instituto de Astrof\'isica de Canarias, The Johns Hopkins University, 
Kavli Institute for the Physics and Mathematics of the Universe (IPMU) / 
University of Tokyo, the Korean Participation Group, Lawrence Berkeley National Laboratory, 
Leibniz Institut f\"ur Astrophysik Potsdam (AIP),  
Max-Planck-Institut f\"ur Astronomie (MPIA Heidelberg), 
Max-Planck-Institut f\"ur Astrophysik (MPA Garching), 
Max-Planck-Institut f\"ur Extraterrestrische Physik (MPE), 
National Astronomical Observatories of China, New Mexico State University, 
New York University, University of Notre Dame, 
Observat\'ario Nacional / MCTI, The Ohio State University, 
Pennsylvania State University, Shanghai Astronomical Observatory, 
United Kingdom Participation Group,
Universidad Nacional Aut\'onoma de M\'exico, University of Arizona, 
University of Colorado Boulder, University of Oxford, University of Portsmouth, 
University of Utah, University of Virginia, University of Washington, University of Wisconsin, 
Vanderbilt University, and Yale University.\\



\begin{thebibliography}{}
\expandafter\ifx\csname natexlab\endcsname\relax\def\natexlab#1{#1}\fi

\bibitem[{{Abazajian} {et~al.}(2009){Abazajian}, {Adelman-McCarthy},
  {Ag{\"u}eros}, {Allam}, {Allende Prieto}, {An}, {Anderson}, {Anderson},
  {Annis}, {Bahcall}, \& et~al.}]{abazajian09}
{Abazajian}, K.~N., {Adelman-McCarthy}, J.~K., {Ag{\"u}eros}, M.~A., {et~al.}
  2009, \apjs, 182, 543

\bibitem[{{Abolfathi} {et~al.}(2018){Abolfathi}, {Aguado}, {Aguilar}, {Allende
  Prieto}, {Almeida}, {Ananna}, {Anders}, {Anderson}, {Andrews}, {Anguiano}, \&
  et~al.}]{masters18}
{Abolfathi}, B., {Aguado}, D.~S., {Aguilar}, G., {et~al.} 2018, \apjs, 235, 42

\bibitem[{{Adelman-McCarthy} {et~al.}(2008){Adelman-McCarthy}, {Ag{\"u}eros},
  {Allam}, {Allende Prieto}, {Anderson}, {Anderson}, {Annis}, {Bahcall},
  {Bailer-Jones}, {Baldry}, {Barentine}, {Bassett}, {Becker}, {Beers}, {Bell},
  {Berlind}, {Bernardi}, {Blanton}, {Bochanski}, {Boroski}, {Brinchmann},
  {Brinkmann}, {Brunner}, {Budav{\'a}ri}, {Carliles}, {Carr}, {Castander},
  {Cinabro}, {Cool}, {Covey}, {Csabai}, {Cunha}, {Davenport}, {Dilday}, {Doi},
  {Eisenstein}, {Evans}, {Fan}, {Finkbeiner}, {Friedman}, {Frieman},
  {Fukugita}, {G{\"a}nsicke}, {Gates}, {Gillespie}, {Glazebrook}, {Gray},
  {Grebel}, {Gunn}, {Gurbani}, {Hall}, {Harding}, {Harvanek}, {Hawley},
  {Hayes}, {Heckman}, {Hendry}, {Hindsley}, {Hirata}, {Hogan}, {Hogg}, {Hyde},
  {Ichikawa}, {Ivezi{\'c}}, {Jester}, {Johnson}, {Jorgensen}, {Juri{\'c}},
  {Kent}, {Kessler}, {Kleinman}, {Knapp}, {Kron}, {Krzesinski}, {Kuropatkin},
  {Lamb}, {Lampeitl}, {Lebedeva}, {Lee}, {French Leger}, {L{\'e}pine}, {Lima},
  {Lin}, {Long}, {Loomis}, {Loveday}, {Lupton}, {Malanushenko}, {Malanushenko},
  {Mandelbaum}, {Margon}, {Marriner}, {Mart{\'{\i}}nez-Delgado}, {Matsubara},
  {McGehee}, {McKay}, {Meiksin}, {Morrison}, {Munn}, {Nakajima}, {Neilsen},
  {Newberg}, {Nichol}, {Nicinski}, {Nieto-Santisteban}, {Nitta}, {Okamura},
  {Owen}, {Oyaizu}, {Padmanabhan}, {Pan}, {Park}, {Peoples}, {Pier}, {Pope},
  {Purger}, {Raddick}, {Re Fiorentin}, {Richards}, {Richmond}, {Riess}, {Rix},
  {Rockosi}, {Sako}, {Schlegel}, {Schneider}, {Schreiber}, {Schwope}, {Seljak},
  {Sesar}, {Sheldon}, {Shimasaku}, {Sivarani}, {Allyn Smith}, {Snedden},
  {Steinmetz}, {Strauss}, {SubbaRao}, {Suto}, {Szalay}, {Szapudi}, {Szkody},
  {Tegmark}, {Thakar}, {Tremonti}, {Tucker}, {Uomoto}, {Vanden Berk},
  {Vandenberg}, {Vidrih}, {Vogeley}, {Voges}, {Vogt}, {Wadadekar}, {Weinberg},
  {West}, {White}, {Wilhite}, {Yanny}, {Yocum}, {York}, {Zehavi}, \&
  {Zucker}}]{adc08}
{Adelman-McCarthy}, J.~K., {Ag{\"u}eros}, M.~A., {Allam}, S.~S., {et~al.} 2008,
  \apjs, 175, 297

\bibitem[{{Albareti} {et~al.}(2017){Albareti}, {Allende Prieto}, {Almeida},
  {Anders}, {Anderson}, {Andrews}, {Arag{\'o}n-Salamanca},
  {Argudo-Fern{\'a}ndez}, {Armengaud}, {Aubourg}, \& et~al.}]{sdss16}
{Albareti}, F.~D., {Allende Prieto}, C., {Almeida}, A., {et~al.} 2017, \apjs,
  233, 25

\bibitem[{{Becker} {et~al.}(1995){Becker}, {White}, \& {Helfand}}]{becker95}
{Becker}, R.~H., {White}, R.~L., \& {Helfand}, D.~J. 1995, \apj, 450, 559

\bibitem[{{Belfiore} {et~al.}(2016){Belfiore}, {Maiolino}, {Maraston},
  {Emsellem}, {Bershady}, {Masters}, {Yan}, {Bizyaev}, {Boquien}, {Brownstein},
  {Bundy}, {Drory}, {Heckman}, {Law}, {Roman-Lopes}, {Pan}, {Stanghellini},
  {Thomas}, {Weijmans}, \& {Westfall}}]{belfiore16}
{Belfiore}, F., {Maiolino}, R., {Maraston}, C., {et~al.} 2016, \mnras, 461,
  3111

\bibitem[{{Belfiore} {et~al.}(2017){Belfiore}, {Maiolino}, {Maraston},
  {Emsellem}, {Bershady}, {Masters}, {Bizyaev}, {Boquien}, {Brownstein},
  {Bundy}, {Diamond-Stanic}, {Drory}, {Heckman}, {Law}, {Malanushenko},
  {Oravetz}, {Pan}, {Roman-Lopes}, {Thomas}, {Weijmans}, {Westfall}, \&
  {Yan}}]{belfiore17}
---. 2017, \mnras, 466, 2570

\bibitem[{{Bell} {et~al.}(2004){Bell}, {Wolf}, {Meisenheimer}, {Rix}, {Borch},
  {Dye}, {Kleinheinrich}, {Wisotzki}, \& {McIntosh}}]{bell04}
{Bell}, E.~F., {Wolf}, C., {Meisenheimer}, K., {et~al.} 2004, \apj, 608, 752

\bibitem[{{Benson} {et~al.}(2003){Benson}, {Bower}, {Frenk}, {Lacey}, {Baugh},
  \& {Cole}}]{benson03}
{Benson}, A.~J., {Bower}, R.~G., {Frenk}, C.~S., {et~al.} 2003, \apj, 599, 38

\bibitem[{{Best} {et~al.}(2006){Best}, {Kaiser}, {Heckman}, \&
  {Kauffmann}}]{best06}
{Best}, P.~N., {Kaiser}, C.~R., {Heckman}, T.~M., \& {Kauffmann}, G. 2006,
  \mnras, 368, L67

\bibitem[{{Best} {et~al.}(2005){Best}, {Kauffmann}, {Heckman}, {Brinchmann},
  {Charlot}, {Ivezi{\'c}}, \& {White}}]{best05}
{Best}, P.~N., {Kauffmann}, G., {Heckman}, T.~M., {et~al.} 2005, \mnras, 362,
  25

\bibitem[{{Binette} {et~al.}(1994){Binette}, {Magris}, {Stasi{\'n}ska}, \&
  {Bruzual}}]{binette94}
{Binette}, L., {Magris}, C.~G., {Stasi{\'n}ska}, G., \& {Bruzual}, A.~G. 1994,
  \aap, 292, 13

\bibitem[{{Binney} \& {Tabor}(1995)}]{binney95}
{Binney}, J., \& {Tabor}, G. 1995, \mnras, 276, 663

\bibitem[{{Blanton} {et~al.}(2003){Blanton}, {Hogg}, {Bahcall}, {Baldry},
  {Brinkmann}, {Csabai}, {Eisenstein}, {Fukugita}, {Gunn}, {Ivezi{\'c}},
  {Lamb}, {Lupton}, {Loveday}, {Munn}, {Nichol}, {Okamura}, {Schlegel},
  {Shimasaku}, {Strauss}, {Vogeley}, \& {Weinberg}}]{blanton03}
{Blanton}, M.~R., {Hogg}, D.~W., {Bahcall}, N.~A., {et~al.} 2003, \apj, 594,
  186

\bibitem[{{Blanton} {et~al.}(2005){Blanton}, {Schlegel}, {Strauss},
  {Brinkmann}, {Finkbeiner}, {Fukugita}, {Gunn}, {Hogg}, {Ivezi{\'c}}, {Knapp},
  {Lupton}, {Munn}, {Schneider}, {Tegmark}, \& {Zehavi}}]{blanton05}
{Blanton}, M.~R., {Schlegel}, D.~J., {Strauss}, M.~A., {et~al.} 2005, \aj, 129,
  2562

\bibitem[{{Blanton} {et~al.}(2017){Blanton}, {Bershady}, {Abolfathi},
  {Albareti}, {Allende Prieto}, {Almeida}, {Alonso-Garc{\'{\i}}a}, {Anders},
  {Anderson}, {Andrews}, \& et~al.}]{blanton17}
{Blanton}, M.~R., {Bershady}, M.~A., {Abolfathi}, B., {et~al.} 2017, \aj, 154,
  28

\bibitem[{{Bower} {et~al.}(2006){Bower}, {Benson}, {Malbon}, {Helly}, {Frenk},
  {Baugh}, {Cole}, \& {Lacey}}]{bower06}
{Bower}, R.~G., {Benson}, A.~J., {Malbon}, R., {et~al.} 2006, \mnras, 370, 645

\bibitem[{{Brown} {et~al.}(2017){Brown}, {Moustakas}, {Kennicutt}, {Bonne},
  {Intema}, {de Gasperin}, {Boquien}, {Jarrett}, {Cluver}, {Smith}, {da Cunha},
  {Imanishi}, {Armus}, {Brandl}, \& {Peek}}]{brown17}
{Brown}, M.~J.~I., {Moustakas}, J., {Kennicutt}, R.~C., {et~al.} 2017, \apj,
  847, 136

\bibitem[{{Bundy} {et~al.}(2006){Bundy}, {Ellis}, {Conselice}, {Taylor},
  {Cooper}, {Willmer}, {Weiner}, {Coil}, {Noeske}, \& {Eisenhardt}}]{bundy06}
{Bundy}, K., {Ellis}, R.~S., {Conselice}, C.~J., {et~al.} 2006, \apj, 651, 120

\bibitem[{{Bundy} {et~al.}(2015){Bundy}, {Bershady}, {Law}, {Yan}, {Drory},
  {MacDonald}, {Wake}, {Cherinka}, {S{\'a}nchez-Gallego}, {Weijmans}, {Thomas},
  {Tremonti}, {Masters}, {Coccato}, {Diamond-Stanic}, {Arag{\'o}n-Salamanca},
  {Avila-Reese}, {Badenes}, {Falc{\'o}n-Barroso}, {Belfiore}, {Bizyaev},
  {Blanc}, {Bland-Hawthorn}, {Blanton}, {Brownstein}, {Byler}, {Cappellari},
  {Conroy}, {Dutton}, {Emsellem}, {Etherington}, {Frinchaboy}, {Fu}, {Gunn},
  {Harding}, {Johnston}, {Kauffmann}, {Kinemuchi}, {Klaene}, {Knapen},
  {Leauthaud}, {Li}, {Lin}, {Maiolino}, {Malanushenko}, {Malanushenko}, {Mao},
  {Maraston}, {McDermid}, {Merrifield}, {Nichol}, {Oravetz}, {Pan}, {Parejko},
  {Sanchez}, {Schlegel}, {Simmons}, {Steele}, {Steinmetz}, {Thanjavur},
  {Thompson}, {Tinker}, {van den Bosch}, {Westfall}, {Wilkinson}, {Wright},
  {Xiao}, \& {Zhang}}]{bundy15}
{Bundy}, K., {Bershady}, M.~A., {Law}, D.~R., {et~al.} 2015, \apj, 798, 7

\bibitem[{{Buson} {et~al.}(1993){Buson}, {Sadler}, {Zeilinger}, {Bertin},
  {Bertola}, {Danzinger}, {Dejonghe}, {Saglia}, \& {de Zeeuw}}]{buson93}
{Buson}, L.~M., {Sadler}, E.~M., {Zeilinger}, W.~W., {et~al.} 1993, \aap, 280,
  409

\bibitem[{{Calzetti} {et~al.}(2000){Calzetti}, {Armus}, {Bohlin}, {Kinney},
  {Koornneef}, \& {Storchi-Bergmann}}]{calzetti00}
{Calzetti}, D., {Armus}, L., {Bohlin}, R.~C., {et~al.} 2000, \apj, 533, 682

\bibitem[{{Cappellari}(2008)}]{cappellari08}
{Cappellari}, M. 2008, \mnras, 390, 71

\bibitem[{{Cappellari}(2017)}]{cappellari17}
---. 2017, \mnras, 466, 798

\bibitem[{{Cappellari} \& {Emsellem}(2004)}]{cappellari04}
{Cappellari}, M., \& {Emsellem}, E. 2004, \pasp, 116, 138

\bibitem[{{Cattaneo} {et~al.}(2009){Cattaneo}, {Faber}, {Binney}, {Dekel},
  {Kormendy}, {Mushotzky}, {Babul}, {Best}, {Br{\"u}ggen}, {Fabian}, {Frenk},
  {Khalatyan}, {Netzer}, {Mahdavi}, {Silk}, {Steinmetz}, \&
  {Wisotzki}}]{cattaneo09}
{Cattaneo}, A., {Faber}, S.~M., {Binney}, J., {et~al.} 2009, \nat, 460, 213

\bibitem[{{Chang} {et~al.}(2015){Chang}, {Ip}, {Lin}, {Cheng}, {Ngeow}, {Yang},
  {Waszczak}, {Kulkarni}, {Levitan}, {Sesar}, {Laher}, {Surace}, \&
  {Prince}}]{chang15}
{Chang}, C.-K., {Ip}, W.-H., {Lin}, H.-W., {et~al.} 2015, \apjs, 219, 27

\bibitem[{{Chen} {et~al.}(2016){Chen}, {Shi}, {Tremonti}, {Bershady},
  {Merrifield}, {Emsellem}, {Jin}, {Huang}, {Fu}, {Wake}, {Bundy}, {Stark},
  {Lin}, {Argudo-Fernandez}, {Bergmann}, {Bizyaev}, {Brownstein}, {Bureau},
  {Chisholm}, {Drory}, {Guo}, {Hao}, {Hu}, {Li}, {Li}, {Roman Lopes}, {Pan},
  {Riffel}, {Thomas}, {Wang}, {Westfall}, \& {Yan}}]{chen16}
{Chen}, Y.-M., {Shi}, Y., {Tremonti}, C.~A., {et~al.} 2016, Nature
  Communications, 7, 13269

\bibitem[{{Cheung} {et~al.}(2016){Cheung}, {Bundy}, {Cappellari}, {Peirani},
  {Rujopakarn}, {Westfall}, {Yan}, {Bershady}, {Greene}, {Heckman}, {Drory},
  {Law}, {Masters}, {Thomas}, {Wake}, {Weijmans}, {Rubin}, {Belfiore},
  {Vulcani}, {Chen}, {Zhang}, {Gelfand}, {Bizyaev}, {Roman-Lopes}, \&
  {Schneider}}]{cheung16}
{Cheung}, E., {Bundy}, K., {Cappellari}, M., {et~al.} 2016, \nat, 533, 504

\bibitem[{{Choi} {et~al.}(2014){Choi}, {Conroy}, {Moustakas}, {Graves},
  {Holden}, {Brodwin}, {Brown}, \& {van Dokkum}}]{choi14}
{Choi}, J., {Conroy}, C., {Moustakas}, J., {et~al.} 2014, \apj, 792, 95

\bibitem[{{Ciotti} \& {Ostriker}(2001)}]{ciotti01}
{Ciotti}, L., \& {Ostriker}, J.~P. 2001, \apj, 551, 131

\bibitem[{{Ciotti} \& {Ostriker}(2007)}]{ciotti07}
---. 2007, \apj, 665, 1038

\bibitem[{{Ciotti} {et~al.}(2010){Ciotti}, {Ostriker}, \& {Proga}}]{ciotti10}
{Ciotti}, L., {Ostriker}, J.~P., \& {Proga}, D. 2010, \apj, 717, 708

\bibitem[{{Condon}(1984)}]{condon84}
{Condon}, J.~J. 1984, \apj, 287, 461

\bibitem[{{Conroy} {et~al.}(2014){Conroy}, {Graves}, \& {van
  Dokkum}}]{conroy14}
{Conroy}, C., {Graves}, G.~J., \& {van Dokkum}, P.~G. 2014, \apj, 780, 33

\bibitem[{{Conroy} {et~al.}(2015){Conroy}, {van Dokkum}, \&
  {Kravtsov}}]{conroy15}
{Conroy}, C., {van Dokkum}, P.~G., \& {Kravtsov}, A. 2015, \apj, 803, 77

\bibitem[{{Croton} {et~al.}(2006){Croton}, {Springel}, {White}, {De Lucia},
  {Frenk}, {Gao}, {Jenkins}, {Kauffmann}, {Navarro}, \& {Yoshida}}]{croton06}
{Croton}, D.~J., {Springel}, V., {White}, S.~D.~M., {et~al.} 2006, \mnras, 365,
  11

\bibitem[{{Demoulin-Ulrich} {et~al.}(1984){Demoulin-Ulrich}, {Butcher}, \&
  {Boksenberg}}]{demoulin84}
{Demoulin-Ulrich}, M.-H., {Butcher}, H.~R., \& {Boksenberg}, A. 1984, \apj,
  285, 527

\bibitem[{{Drory} {et~al.}(2015){Drory}, {MacDonald}, {Bershady}, {Bundy},
  {Gunn}, {Law}, {Smith}, {Stoll}, {Tremonti}, {Wake}, {Yan}, {Weijmans},
  {Byler}, {Cherinka}, {Cope}, {Eigenbrot}, {Harding}, {Holder}, {Huehnerhoff},
  {Jaehnig}, {Jansen}, {Klaene}, {Paat}, {Percival}, \& {Sayres}}]{drory15}
{Drory}, N., {MacDonald}, N., {Bershady}, M.~A., {et~al.} 2015, \aj, 149, 77

\bibitem[{{Dunlop} \& {Peacock}(1990)}]{dunlop90}
{Dunlop}, J.~S., \& {Peacock}, J.~A. 1990, \mnras, 247, 19

\bibitem[{{Dunn} \& {Fabian}(2006)}]{dunn06}
{Dunn}, R.~J.~H., \& {Fabian}, A.~C. 2006, \mnras, 373, 959

\bibitem[{{Faber} {et~al.}(2007){Faber}, {Willmer}, {Wolf}, {Koo}, {Weiner},
  {Newman}, {Im}, {Coil}, {Conroy}, {Cooper}, {Davis}, {Finkbeiner}, {Gerke},
  {Gebhardt}, {Groth}, {Guhathakurta}, {Harker}, {Kaiser}, {Kassin},
  {Kleinheinrich}, {Konidaris}, {Kron}, {Lin}, {Luppino}, {Madgwick},
  {Meisenheimer}, {Noeske}, {Phillips}, {Sarajedini}, {Schiavon}, {Simard},
  {Szalay}, {Vogt}, \& {Yan}}]{faber07}
{Faber}, S.~M., {Willmer}, C.~N.~A., {Wolf}, C., {et~al.} 2007, \apj, 665, 265

\bibitem[{{Fabian}(1994)}]{fabian94}
{Fabian}, A.~C. 1994, \araa, 32, 277

\bibitem[{{Fabian}(2012)}]{fabian12}
---. 2012, \araa, 50, 455

\bibitem[{{Fabian} {et~al.}(2006){Fabian}, {Sanders}, {Taylor}, {Allen},
  {Crawford}, {Johnstone}, \& {Iwasawa}}]{fabian06}
{Fabian}, A.~C., {Sanders}, J.~S., {Taylor}, G.~B., {et~al.} 2006, \mnras, 366,
  417

\bibitem[{{Falcke} \& {Biermann}(1999)}]{falcke99}
{Falcke}, H., \& {Biermann}, P.~L. 1999, \aap, 342, 49

\bibitem[{{Falc{\'o}n-Barroso} {et~al.}(2011){Falc{\'o}n-Barroso},
  {S{\'a}nchez-Bl{\'a}zquez}, {Vazdekis}, {Ricciardelli}, {Cardiel}, {Cenarro},
  {Gorgas}, \& {Peletier}}]{MILES}
{Falc{\'o}n-Barroso}, J., {S{\'a}nchez-Bl{\'a}zquez}, P., {Vazdekis}, A.,
  {et~al.} 2011, \aap, 532, A95

\bibitem[{{Graham} {et~al.}(2018){Graham}, {Cappellari}, {Li}, {Mao},
  {Bershady}, {Bizyaev}, {Brinkmann}, {Brownstein}, {Bundy}, {Drory}, {Law},
  {Pan}, {Thomas}, {Wake}, {Weijmans}, {Westfall}, \& {Yan}}]{graham18}
{Graham}, M.~T., {Cappellari}, M., {Li}, H., {et~al.} 2018, \mnras, 477, 4711

\bibitem[{{Graves} \& {Schiavon}(2008)}]{graves08}
{Graves}, G.~J., \& {Schiavon}, R.~P. 2008, \apjs, 177, 446

\bibitem[{{Gunn} {et~al.}(2006){Gunn}, {Siegmund}, {Mannery}, {Owen}, {Hull},
  {Leger}, {Carey}, {Knapp}, {York}, {Boroski}, {Kent}, {Lupton}, {Rockosi},
  {Evans}, {Waddell}, {Anderson}, {Annis}, {Barentine}, {Bartoszek}, {Bastian},
  {Bracker}, {Brewington}, {Briegel}, {Brinkmann}, {Brown}, {Carr},
  {Czarapata}, {Drennan}, {Dombeck}, {Federwitz}, {Gillespie}, {Gonzales},
  {Hansen}, {Harvanek}, {Hayes}, {Jordan}, {Kinney}, {Klaene}, {Kleinman},
  {Kron}, {Kresinski}, {Lee}, {Limmongkol}, {Lindenmeyer}, {Long}, {Loomis},
  {McGehee}, {Mantsch}, {Neilsen}, {Neswold}, {Newman}, {Nitta}, {Peoples},
  {Pier}, {Prieto}, {Prosapio}, {Rivetta}, {Schneider}, {Snedden}, \&
  {Wang}}]{gunn06}
{Gunn}, J.~E., {Siegmund}, W.~A., {Mannery}, E.~J., {et~al.} 2006, \aj, 131,
  2332

\bibitem[{{Heckman} \& {Best}(2014)}]{heckman14}
{Heckman}, T.~M., \& {Best}, P.~N. 2014, \araa, 52, 589

\bibitem[{{Ilbert} {et~al.}(2010){Ilbert}, {Salvato}, {Le Floc'h}, {Aussel},
  {Capak}, {McCracken}, {Mobasher}, {Kartaltepe}, {Scoville}, {Sanders},
  {Arnouts}, {Bundy}, {Cassata}, {Kneib}, {Koekemoer}, {Le F{\`e}vre}, {Lilly},
  {Surace}, {Taniguchi}, {Tasca}, {Thompson}, {Tresse}, {Zamojski}, {Zamorani},
  \& {Zucca}}]{ilbert10}
{Ilbert}, O., {Salvato}, M., {Le Floc'h}, E., {et~al.} 2010, \apj, 709, 644

\bibitem[{{Jiang} {et~al.}(2010){Jiang}, {Ciotti}, {Ostriker}, \&
  {Spitkovsky}}]{jiang10}
{Jiang}, Y.-F., {Ciotti}, L., {Ostriker}, J.~P., \& {Spitkovsky}, A. 2010,
  \apj, 711, 125

\bibitem[{{Kauffmann} {et~al.}(2003){Kauffmann}, {Heckman}, {White}, {Charlot},
  {Tremonti}, {Brinchmann}, {Bruzual}, {Peng}, {Seibert}, {Bernardi},
  {Blanton}, {Brinkmann}, {Castander}, {Cs{\'a}bai}, {Fukugita}, {Ivezic},
  {Munn}, {Nichol}, {Padmanabhan}, {Thakar}, {Weinberg}, \&
  {York}}]{kauffman03}
{Kauffmann}, G., {Heckman}, T.~M., {White}, S.~D.~M., {et~al.} 2003, \mnras,
  341, 33

\bibitem[{{Kennicutt} {et~al.}(2009){Kennicutt}, {Hao}, {Calzetti},
  {Moustakas}, {Dale}, {Bendo}, {Engelbracht}, {Johnson}, \&
  {Lee}}]{kennicutt09}
{Kennicutt}, Jr., R.~C., {Hao}, C.-N., {Calzetti}, D., {et~al.} 2009, \apj,
  703, 1672

\bibitem[{{Kroupa} \& {Weidner}(2003)}]{kroupa}
{Kroupa}, P., \& {Weidner}, C. 2003, \apj, 598, 1076

\bibitem[{{Lagos} {et~al.}(2015){Lagos}, {Crain}, {Schaye}, {Furlong}, {Frenk},
  {Bower}, {Schaller}, {Theuns}, {Trayford}, {Bah{\'e}}, \& {Dalla
  Vecchia}}]{lagos15}
{Lagos}, C.~d.~P., {Crain}, R.~A., {Schaye}, J., {et~al.} 2015, \mnras, 452,
  3815

\bibitem[{{Law} {et~al.}(2015){Law}, {Yan}, {Bershady}, {Bundy}, {Cherinka},
  {Drory}, {MacDonald}, {S{\'a}nchez-Gallego}, {Wake}, {Weijmans}, {Blanton},
  {Klaene}, {Moran}, {Sanchez}, \& {Zhang}}]{law15}
{Law}, D.~R., {Yan}, R., {Bershady}, M.~A., {et~al.} 2015, \aj, 150, 19

\bibitem[{{Law} {et~al.}(2016){Law}, {Cherinka}, {Yan}, {Andrews}, {Bershady},
  {Bizyaev}, {Blanc}, {Blanton}, {Bolton}, {Brownstein}, {Bundy}, {Chen},
  {Drory}, {D'Souza}, {Fu}, {Jones}, {Kauffmann}, {MacDonald}, {Masters},
  {Newman}, {Parejko}, {S{\'a}nchez-Gallego}, {S{\'a}nchez}, {Schlegel},
  {Thomas}, {Wake}, {Weijmans}, {Westfall}, \& {Zhang}}]{law16}
{Law}, D.~R., {Cherinka}, B., {Yan}, R., {et~al.} 2016, \aj, 152, 83

\bibitem[{{Martig} {et~al.}(2009){Martig}, {Bournaud}, {Teyssier}, \&
  {Dekel}}]{martig09}
{Martig}, M., {Bournaud}, F., {Teyssier}, R., \& {Dekel}, A. 2009, \apj, 707,
  250

\bibitem[{{Martin} {et~al.}(2005){Martin}, {Seibert}, {Buat},
  {Iglesias-P{\'a}ramo}, {Barlow}, {Bianchi}, {Byun}, {Donas}, {Forster},
  {Friedman}, {Heckman}, {Jelinsky}, {Lee}, {Madore}, {Malina}, {Milliard},
  {Morrissey}, {Neff}, {Rich}, {Schiminovich}, {Siegmund}, {Small}, {Szalay},
  {Welsh}, \& {Wyder}}]{martin05}
{Martin}, D.~C., {Seibert}, M., {Buat}, V., {et~al.} 2005, \apjl, 619, L59

\bibitem[{{Mathews} \& {Brighenti}(2003)}]{mathews03}
{Mathews}, W.~G., \& {Brighenti}, F. 2003, \araa, 41, 191

\bibitem[{{McNamara} \& {Nulsen}(2007)}]{mcnamara07}
{McNamara}, B.~R., \& {Nulsen}, P.~E.~J. 2007, \araa, 45, 117

\bibitem[{{Morganti}(2017)}]{morganti17}
{Morganti}, R. 2017, Frontiers in Astronomy and Space Sciences, 4, 42

\bibitem[{{Moustakas} {et~al.}(2013){Moustakas}, {Coil}, {Aird}, {Blanton},
  {Cool}, {Eisenstein}, {Mendez}, {Wong}, {Zhu}, \& {Arnouts}}]{moustakas13}
{Moustakas}, J., {Coil}, A.~L., {Aird}, J., {et~al.} 2013, \apj, 767, 50

\bibitem[{{Narayan} {et~al.}(2000){Narayan}, {Igumenshchev}, \&
  {Abramowicz}}]{narayan00}
{Narayan}, R., {Igumenshchev}, I.~V., \& {Abramowicz}, M.~A. 2000, \apj, 539,
  798

\bibitem[{{Narayan} {et~al.}(1995){Narayan}, {Yi}, \& {Mahadevan}}]{narayan95}
{Narayan}, R., {Yi}, I., \& {Mahadevan}, R. 1995, \nat, 374, 623

\bibitem[{{O'Sullivan} {et~al.}(2001){O'Sullivan}, {Forbes}, \&
  {Ponman}}]{sullivan01}
{O'Sullivan}, E., {Forbes}, D.~A., \& {Ponman}, T.~J. 2001, \mnras, 328, 461

\bibitem[{{Padmanabhan} {et~al.}(2008){Padmanabhan}, {Schlegel}, {Finkbeiner},
  {Barentine}, {Blanton}, {Brewington}, {Gunn}, {Harvanek}, {Hogg},
  {Ivezi{\'c}}, {Johnston}, {Kent}, {Kleinman}, {Knapp}, {Krzesinski}, {Long},
  {Neilsen}, {Nitta}, {Loomis}, {Lupton}, {Roweis}, {Snedden}, {Strauss}, \&
  {Tucker}}]{padmanabhan08}
{Padmanabhan}, N., {Schlegel}, D.~J., {Finkbeiner}, D.~P., {et~al.} 2008, \apj,
  674, 1217

\bibitem[{{Salim} {et~al.}(2005){Salim}, {Charlot}, {Rich}, {Kauffmann},
  {Heckman}, {Barlow}, {Bianchi}, {Byun}, {Donas}, {Forster}, {Friedman},
  {Jelinsky}, {Lee}, {Madore}, {Malina}, {Martin}, {Milliard}, {Morrissey},
  {Neff}, {Schiminovich}, {Seibert}, {Siegmund}, {Small}, {Szalay}, {Welsh}, \&
  {Wyder}}]{salim05}
{Salim}, S., {Charlot}, S., {Rich}, R.~M., {et~al.} 2005, \apjl, 619, L39

\bibitem[{{Salim} {et~al.}(2007){Salim}, {Rich}, {Charlot}, {Brinchmann},
  {Johnson}, {Schiminovich}, {Seibert}, {Mallery}, {Heckman}, {Forster},
  {Friedman}, {Martin}, {Morrissey}, {Neff}, {Small}, {Wyder}, {Bianchi},
  {Donas}, {Lee}, {Madore}, {Milliard}, {Szalay}, {Welsh}, \& {Yi}}]{salim07}
{Salim}, S., {Rich}, R.~M., {Charlot}, S., {et~al.} 2007, \apjs, 173, 267

\bibitem[{{Salim} {et~al.}(2009){Salim}, {Dickinson}, {Michael Rich},
  {Charlot}, {Lee}, {Schiminovich}, {P{\'e}rez-Gonz{\'a}lez}, {Ashby},
  {Papovich}, {Faber}, {Ivison}, {Frayer}, {Walton}, {Weiner}, {Chary},
  {Bundy}, {Noeske}, \& {Koekemoer}}]{salim09}
{Salim}, S., {Dickinson}, M., {Michael Rich}, R., {et~al.} 2009, \apj, 700, 161

\bibitem[{{Smee} {et~al.}(2013){Smee}, {Gunn}, {Uomoto}, {Roe}, {Schlegel},
  {Rockosi}, {Carr}, {Leger}, {Dawson}, {Olmstead}, {Brinkmann}, {Owen},
  {Barkhouser}, {Honscheid}, {Harding}, {Long}, {Lupton}, {Loomis}, {Anderson},
  {Annis}, {Bernardi}, {Bhardwaj}, {Bizyaev}, {Bolton}, {Brewington}, {Briggs},
  {Burles}, {Burns}, {Castander}, {Connolly}, {Davenport}, {Ebelke}, {Epps},
  {Feldman}, {Friedman}, {Frieman}, {Heckman}, {Hull}, {Knapp}, {Lawrence},
  {Loveday}, {Mannery}, {Malanushenko}, {Malanushenko}, {Merrelli}, {Muna},
  {Newman}, {Nichol}, {Oravetz}, {Pan}, {Pope}, {Ricketts}, {Shelden},
  {Sandford}, {Siegmund}, {Simmons}, {Smith}, {Snedden}, {Schneider},
  {SubbaRao}, {Tremonti}, {Waddell}, \& {York}}]{smee13}
{Smee}, S.~A., {Gunn}, J.~E., {Uomoto}, A., {et~al.} 2013, \aj, 146, 32

\bibitem[{{Strateva} {et~al.}(2001){Strateva}, {Ivezi{\'c}}, {Knapp},
  {Narayanan}, {Strauss}, {Gunn}, {Lupton}, {Schlegel}, {Bahcall}, {Brinkmann},
  {Brunner}, {Budav{\'a}ri}, {Csabai}, {Castander}, {Doi}, {Fukugita}, {Gy{\H
  o}ry}, {Hamabe}, {Hennessy}, {Ichikawa}, {Kunszt}, {Lamb}, {McKay},
  {Okamura}, {Racusin}, {Sekiguchi}, {Schneider}, {Shimasaku}, \&
  {York}}]{strateva01}
{Strateva}, I., {Ivezi{\'c}}, {\v Z}., {Knapp}, G.~R., {et~al.} 2001, \aj, 122,
  1861

\bibitem[{{Thomas} {et~al.}(2005){Thomas}, {Maraston}, {Bender}, \& {Mendes de
  Oliveira}}]{thomas05}
{Thomas}, D., {Maraston}, C., {Bender}, R., \& {Mendes de Oliveira}, C. 2005,
  \apj, 621, 673

\bibitem[{{Tinsley}(1979)}]{tinsley79}
{Tinsley}, B.~M. 1979, \apj, 229, 1046

\bibitem[{{Trager} {et~al.}(2000){Trager}, {Faber}, {Worthey}, \&
  {Gonz{\'a}lez}}]{trager00}
{Trager}, S.~C., {Faber}, S.~M., {Worthey}, G., \& {Gonz{\'a}lez}, J.~J. 2000,
  \aj, 120, 165

\bibitem[{{Wake} {et~al.}(2017){Wake}, {Bundy}, {Diamond-Stanic}, {Yan},
  {Blanton}, {Bershady}, {S{\'a}nchez-Gallego}, {Drory}, {Jones}, {Kauffmann},
  {Law}, {Li}, {MacDonald}, {Masters}, {Thomas}, {Tinker}, {Weijmans}, \&
  {Brownstein}}]{wake17}
{Wake}, D.~A., {Bundy}, K., {Diamond-Stanic}, A.~M., {et~al.} 2017, \aj, 154,
  86

\bibitem[{{Worthey} {et~al.}(1992){Worthey}, {Faber}, \&
  {Gonzalez}}]{worthey92}
{Worthey}, G., {Faber}, S.~M., \& {Gonzalez}, J.~J. 1992, \apj, 398, 69

\bibitem[{{Worthey} {et~al.}(2014){Worthey}, {Tang}, \& {Serven}}]{worthey14}
{Worthey}, G., {Tang}, B., \& {Serven}, J. 2014, \apj, 783, 20

\bibitem[{{Yan} {et~al.}(2016){Yan}, {Bundy}, {Law}, {Bershady}, {Andrews},
  {Cherinka}, {Diamond-Stanic}, {Drory}, {MacDonald}, {S{\'a}nchez-Gallego},
  {Thomas}, {Wake}, {Weijmans}, {Westfall}, {Zhang}, {Arag{\'o}n-Salamanca},
  {Belfiore}, {Bizyaev}, {Blanc}, {Blanton}, {Brownstein}, {Cappellari},
  {D'Souza}, {Emsellem}, {Fu}, {Gaulme}, {Graham}, {Goddard}, {Gunn},
  {Harding}, {Jones}, {Kinemuchi}, {Li}, {Li}, {Maiolino}, {Mao}, {Maraston},
  {Masters}, {Merrifield}, {Oravetz}, {Pan}, {Parejko}, {Sanchez}, {Schlegel},
  {Simmons}, {Thanjavur}, {Tinker}, {Tremonti}, {van den Bosch}, \&
  {Zheng}}]{yan16}
{Yan}, R., {Bundy}, K., {Law}, D.~R., {et~al.} 2016, \aj, 152, 197

\bibitem[{{Yan} {et~al.}(2013){Yan}, {Pan}, {Liu}, {Qu}, {Xue}, {Deng}, {Ma},
  \& {Kong}}]{yan13}
{Yan}, X.~L., {Pan}, G.~M., {Liu}, J.~H., {et~al.} 2013, \aj, 145, 153

\bibitem[{{York} {et~al.}(2000){York}, {Adelman}, {Anderson}, {Anderson},
  {Annis}, {Bahcall}, {Bakken}, {Barkhouser}, {Bastian}, {Berman}, {Boroski},
  {Bracker}, {Briegel}, {Briggs}, {Brinkmann}, {Brunner}, {Burles}, {Carey},
  {Carr}, {Castander}, {Chen}, {Colestock}, {Connolly}, {Crocker}, {Csabai},
  {Czarapata}, {Davis}, {Doi}, {Dombeck}, {Eisenstein}, {Ellman}, {Elms},
  {Evans}, {Fan}, {Federwitz}, {Fiscelli}, {Friedman}, {Frieman}, {Fukugita},
  {Gillespie}, {Gunn}, {Gurbani}, {de Haas}, {Haldeman}, {Harris}, {Hayes},
  {Heckman}, {Hennessy}, {Hindsley}, {Holm}, {Holmgren}, {Huang}, {Hull},
  {Husby}, {Ichikawa}, {Ichikawa}, {Ivezi{\'c}}, {Kent}, {Kim}, {Kinney},
  {Klaene}, {Kleinman}, {Kleinman}, {Knapp}, {Korienek}, {Kron}, {Kunszt},
  {Lamb}, {Lee}, {Leger}, {Limmongkol}, {Lindenmeyer}, {Long}, {Loomis},
  {Loveday}, {Lucinio}, {Lupton}, {MacKinnon}, {Mannery}, {Mantsch}, {Margon},
  {McGehee}, {McKay}, {Meiksin}, {Merelli}, {Monet}, {Munn}, {Narayanan},
  {Nash}, {Neilsen}, {Neswold}, {Newberg}, {Nichol}, {Nicinski}, {Nonino},
  {Okada}, {Okamura}, {Ostriker}, {Owen}, {Pauls}, {Peoples}, {Peterson},
  {Petravick}, {Pier}, {Pope}, {Pordes}, {Prosapio}, {Rechenmacher}, {Quinn},
  {Richards}, {Richmond}, {Rivetta}, {Rockosi}, {Ruthmansdorfer}, {Sandford},
  {Schlegel}, {Schneider}, {Sekiguchi}, {Sergey}, {Shimasaku}, {Siegmund},
  {Smee}, {Smith}, {Snedden}, {Stone}, {Stoughton}, {Strauss}, {Stubbs},
  {SubbaRao}, {Szalay}, {Szapudi}, {Szokoly}, {Thakar}, {Tremonti}, {Tucker},
  {Uomoto}, {Vanden Berk}, {Vogeley}, {Waddell}, {Wang}, {Watanabe},
  {Weinberg}, {Yanny}, {Yasuda}, \& {SDSS Collaboration}}]{york00}
{York}, D.~G., {Adelman}, J., {Anderson}, Jr., J.~E., {et~al.} 2000, \aj, 120,
  1579

\bibitem[{{Yuan} \& {Narayan}(2014)}]{yuan14}
{Yuan}, F., \& {Narayan}, R. 2014, \araa, 52, 529

\bibitem[{{Zensus}(1997)}]{zensus97}
{Zensus}, J.~A. 1997, \araa, 35, 607

\end{thebibliography}
\end{document}